\documentclass[3p, 11pt]{elsarticle}
\makeatletter
\def\ps@pprintTitle{%
  \let\@oddhead\@empty
  \let\@evenhead\@empty
  \def\@oddfoot{\empty}
  \def\@evenfoot{\empty}
}
\makeatother
\usepackage{amssymb}
\usepackage{amsmath}
\usepackage{mathrsfs}
\usepackage{hyperref}
\usepackage{cleveref}
\usepackage[utf8]{inputenc}
\usepackage[T1]{fontenc}
\usepackage{caption}
\usepackage{subcaption}
\usepackage{tabularx}
\usepackage{amsmath}
\usepackage{lineno}

\usepackage{natbib}  
\usepackage{xcolor}
\usepackage{listings}
\newcommand{\non}{\nonumber}
\newcommand{\ov}{\overline}

\newcommand{\la}{\left<}
\newcommand{\ra}{\right>}
\newcommand{\lp}{\left(}
\newcommand{\rp}{\right)}

\newcommand{\bx}{\mbox{${\bf x}$}}

\newcommand{\rhob}{\mbox{$\la \rho \ra_{\ell}$}}

\newcommand{\uib}{\mbox{$\la u_i \ra_L$}}

\newcommand{\taub}{\mbox{${\tau}$}}

\newcommand{\bphi}{\mbox{\boldmath $\phi$}}
\newcommand{\bpsi}{\mbox{\boldmath $\psi$}}

\newcommand{\Phat}{\hat{\Phi}}

\newtheorem{definition}{Definition}

\begin{document}

\begin{frontmatter}

\title{On-the-fly Reduced-Order Modeling of the Filter Density Function with Time-Dependent Subspaces}

\author[inst2]{Aidyn Aitzhan}
\affiliation[inst2]{NVIDIA Corporation, Santa Clara, CA, 95051, USA}
\affiliation[inst1]{ 
 Mechanical Engineering and Materials Science, University of Pittsburgh, Pittsburgh, PA, 15261, USA}%
\author[inst1]{Peyman Givi}
\author[inst1]{Hessam Babaee
\corref{mycorrespondingauthor}}
\ead{h.babaee@pitt.edu}
\cortext[mycorrespondingauthor]{Corresponding author}

\begin{abstract}

A dynamical low-rank approximation is developed for reduced-order modeling (ROM) of the filtered density function (FDF) transport equation, which is utilized for large eddy simulation (LES) of turbulent reacting flows.  In this methodology, the evolution of the composition matrix describing the FDF transport via a set of Langevin equations is constrained to a low-rank matrix manifold.  The composition matrix is approximated using a low-rank factorization, which consists of two thin, time-dependent matrices representing spatial and composition bases, along with a small time-dependent coefficient matrix. The evolution equations for spatial and composition subspaces are derived by projecting the composition transport equation onto the tangent space of the low-rank matrix manifold.  Unlike conventional ROMs, such as those based on principal component analysis, both subspaces are time-dependent and the ROM does not require any prior data to extract the low-dimensional subspaces.  As a result, the constructed ROM adapts on the fly to changes in the dynamics.  For demonstration, LES via the time-dependent bases (TDB)  is conducted of the canonical configuration of a temporally developing planar CO/H\textsubscript{2} jet flame. The flame is rich with strong flame-turbulence interactions resulting in local extinction followed by re-ignition.  The combustion chemistry is modeled via the skeletal kinetics, containing 11 species with 21 reaction steps.  It is shown that the FDF-TDB yields excellent predictions of various statistics of the thermo-chemistry variables, as compared to the full-order model (FOM).
\end{abstract}

\begin{keyword}
time-dependent subspaces; reduced-order modeling; turbulent combustion, LES, FDF
\end{keyword}

\end{frontmatter}

\parindent 0pt
\parskip 0.1in

\section{\label{sec:Introduction}Introduction}

The filtered density function (FDF) \cite{POPE00,Haworth10,Yang2021,Givi06} has proven very effective for large eddy simulation (LES) of turbulent reacting flows. This is due to the inherent capability of the FDF to account for full statistics of the subgrid-scale (SGS) quantities.  The last decade has witnessed a significant increase in fine-tuning and extensive applications of FDF for numerical simulations of a variety of problems in turbulent combustion. See Ref.\ \cite{ZGR2025} for a survey  of the most recent contributions.  Despite its superior performance and popularity, the computational requirements for LES-FDF are relatively high due to the high dimensionality of the FDF when a large number of species are considered \cite{Pope13}.   This is particularly the case when considering the computational memory requirements \cite{PYSG13,ASGG15,AGSG11}. 

Reduced-order modeling has been envisioned as a means of reducing the computational cost of LES-FDF \cite{Pope13}. State-of-the-art approaches to develop reduced order models (ROMs) to alleviate such requirements typically operate in an offline manner in which unimportant reactions of the detailed model are removed. In turbulent flows, these processes are strongly coupled to the hydrodynamics, and an offline reduction strategy may remove reactions that otherwise may affect the flow  (and vice versa). Although the eliminated reactions may only be active for a short time, they can influence the system dynamics by initiating a different reaction chain or triggering nonlinear flow instabilities. Current approaches also require expensive \textit{a priori} simulations to compute the physical sensitivities as required for deriving reduced models. The majority of these techniques are based on static subspaces or manifolds by which all other transport variables can be calculated. An example is the popular principal component analysis (PCA) \cite{SP09,OE17}. However, to the best of our knowledge, PCA-based ROMs have not been developed in the context of FDF.

This work aims to develop an on-the-fly reduced-order model (ROM) based on time-dependent bases (TDBs) to solve the LES-FDF transport equation. Unlike other ROM strategies, the TDB-ROM employs time-evolving subspaces rather than static ones. Furthermore, this model is not \emph{data-driven}, but rather \emph{model-driven}, eliminating the need for an offline stage to extract low-dimensional subspaces from simulation or experimental data. Instead, it provides closed-form equations for both the evolution of the TDBs and the projection of the full-order model (FOM) onto these bases.  The TDB-ROM, as formulated here, is inspired by low-rank approximations that originated in quantum chemistry, specifically the multiconfiguration time-dependent Hartree method (MCTDH) \cite{Beck:2000aa}, which has proven to be highly effective for solving the time-dependent Schrödinger equation. The dynamical low-rank approximation \cite{KL07} extends MCTDH to reduced-order modeling (ROM) of generic matrix differential equations (MDEs). Similar low-rank methods have been developed for solving stochastic partial differential equations \cite{SL09,Babaee:2017aa,MN18,B19,CHZI13,PB20}. Recently, TDB-ROMs have been successfully developed for solving scalar transport in turbulent combustion \cite{RNB21,AITZHAN2025} and constructing skeletal kinetic models in hydrocarbon combustion \cite{NBGCL21,LB24} and astrophysical reaction networks \cite{NLGBL24}.

The novelty of this work is the consideration of the  Langevin equations of the  compositional transport as a matrix differential equation (MDE), and the development of an on-the-fly ROM-TDB to exploit the instantaneous low-dimensional structures of the resulting MDE. The performance of the TDB-ROM is assessed via  LES of the canonical configuration of a temporally developing planar CO/H\textsubscript{2} jet flame. The model predictions are appraised via comparisons against those via the full-order model (FOM).

\section{Methodology}
\subsection{Problem Setup}

For computational description of a low-speed turbulent reacting flow involving $n_s$ species, the primary transport variables are the fluid density $\rho({x}, t) $, the velocity vector $u_i({x}, t),\
i=1,2,3$ along the $x_i$ direction, the total specific enthalpy $h ({x}, t)$,
the pressure $p({x}, t)$, and the species mass fractions $Y_{\alpha}({x}, t) \
(\alpha=1,2,\dots,n_s)$, where $n_s$ denotes the number of species. The conservation equations governing these
variables are the continuity, momentum, enthalpy (energy) and species
mass fraction equations, along with an equation of state. With the low Mach number approximation, the chemical source terms
($S_{\alpha}= S_{\alpha}(\Phi),\ \Phi=[Y_1,Y_2,\dots,Y_{n_s}, h],\ \alpha=1,2,,\dots n_c=n_s+1$) are functions of the composition variables ($\bphi$) only. Thus $n_c$ denotes the total number of scalar variables considered in FDF. For LES, all the transport parameters, say $Q(\bx,t)$ are spatially averaged by passing through a filter of characteristic width $\Delta_G$. The filtered variable is denoted by 
$ \la Q(\bx,t) \ra_{\ell}$, and 
$ {\la Q(\bx,t) \ra_L =} {\la \rho Q \ra_{\ell}} / {\rhob } $ represents its density weighted (Favre) average. Considering the statistics of only the scalar variables (mass fractions of the species and the total enthalpy), the FDF is denoted by $ F_L\lp\bpsi; \bx,t\rp$, where $\bpsi$ represented the entire probability domain for the scalars field. In the modeled equations representing transport of FDF, the effects of the subgrid-scale (SGS) convection are modeled by the standard gradient diffusion model \cite{Smagorinsky63,garnier2009large} with the Vreman's model \cite{Vreman2004Eddy} for the SGS viscosity, and unity SGS Prandtl and Schmidt numbers. The influence of SGS mixing is taken into account with the LMSE/IEM closure \cite{OBrien80}. With these models, the FDF transport equation is of the form \cite{ASGG15}:
\begin{align}
\frac{\partial F_L}{\partial t} + \frac{\partial [ \uib F_L]}{\partial
x_i} &= \frac{\partial}{\partial x_i} \left[(\gamma + \gamma_t)
\frac{\partial (F_L / \rhob)}
{\partial x_i}\right] \non \\ 
&+ \frac{\partial}{\partial
\psi_{\alpha} }\left[ \Omega (\psi_{\alpha} -
\la \phi_{\alpha} \ra_{L}) F_L \right] - \frac{\partial}{\partial
\psi_{\alpha} }\left[ S_{\alpha}\left( \bpsi \right) F_L \right],
\label{EQ:29}
\end{align}
where $\gamma$, and $\gamma_t$ denote the molecular and the SGS diffusivity, respectively \cite{Vreman2004Eddy}. The term $\Omega =C_{\phi}\lp \gamma +\gamma_{t}\rp / \lp \la \rho \ra_{\ell}\Delta_G^{2} \rp$ is the modeled SGS mixing frequency \cite{MLP03,OBrien80} with the  model constant $C_\phi$. All of these model parameters are standard, and are the same as those employed in previous work; e.g.\ Ref.\ \cite{Aitzhan2022CTM}.   Equation (\ref{EQ:29}) may be integrated to obtain the modeled transport equations for the SGS moments, e.g. the filtered mean, $\la \phi_{k} \ra_{L}$ and the SGS variance $\taub_{k} \equiv \la \phi_{k}^2 \ra_L- \la \phi_{k} \ra^2_L$. A convenient means of solving this equation is via the Lagrangian Monte Carlo (MC) procedure \cite{Hiremath2012,MJG06,Jaberi2020}. In this procedure, each of the MC elements (particles) undergoes motion in physical space by convection due to the filtered mean flow velocity and diffusion due to molecular and subgrid diffusivities. These are determined by viewing Eq.\ (\ref{EQ:29}) as a Fokker-Planck equation, for which the corresponding Langevin equations describing the transport of the MC particles are \cite{Risken89,Gardiner90}: 
\begin{equation}
dX_i(t)=\left[\uib + \frac{1}{\rhob} \frac{\partial (\gamma+\gamma_t)}
{\partial x_i} \right] dt+\sqrt{ 2(\gamma+\gamma_t) / \rhob }\ dW_i(t),
\label{EQ:MC1}
\end{equation}
with the change in the compositional makeup according to: 
\begin{equation}
\frac{d \phi_{k}^{+}}{dt} = -\Omega \left( \phi_{k}^{+} - \la \phi_{k}\ra_{L} \right) + S_{k}\left( \bphi^{+} \right) \; \left(k = 1,2, \dots n_s+1 \right).
\label{EQ:MC2}
\end{equation}
In these equations, ${W}_i$ denotes the Wiener-Levy process, $\phi_{k}^{+} = \phi_{k}\left(\mathbf X, t\right)$ is the scalar value of the particle with the Lagrangian position $X_i$. 

\subsection{Reduced-Order Modeling via Time-Dependent Bases}
A key contribution of this work is to recast Eq.\ (\ref{EQ:MC2})  as an MDE as follows:
\begin{equation}\label{eq:FOM}
\frac{d\Phi}{dt} =M (\Phi) = - L(\Phi) + S(\Phi),
\end{equation}
where $\Phi(t) \in \mathbb{R}^{n \times n_c}$ is the composition matrix:
\begin{align}
\Phi(t) =[\phi_1(t) &\mid \phi_2(t) \mid \dots \mid \phi_{n_s}(t) \mid \phi_{n_s+1}(t)],
\end{align}
with $n_c$ columns $(n_c=n_s+1)$, and $n$ particles and in which $\phi_{n_c} \equiv h$. Therefore, the $i$-th row of $\Phi$ contains the composition values of the $i$-th particle. The right-hand side of Eq.\ (\ref{eq:FOM}) contains two terms: (i) $S(\Phi)$ is the matrix of chemical and energy source terms such that 
$S(\Phi) = [S(\phi_1) | S(\phi_2) | \dots S(\phi_{n_s}) | S(\phi_{n_s+1}) ] $ denotes the chemical source terms.
 Similar to $\Phi$, the $i$-th row of $S(\Phi)$ contains the chemical source term for the $i$-th particle. (ii) $L(\Phi)=\Omega \left( \Phi - \la \Phi \ra_{L} \right): \mathbb{R}^{n \times n_c} \rightarrow   \mathbb{R}^{n \times n_c}$ is the mixing term.  
Through numerical demonstration, it is shown that the matrix $\Phi(t)$ allows for accurate instantaneous low-rank approximations. Therefore, a low-rank approximation is formulated based on TDB to exploit these structures, with an accurate approximation of the  MDE with significantly fewer degrees of freedom. These low-rank structures, however, do not emerge when Eq.\ (\ref{EQ:MC2}) is solved as a vector differential equation, as done conventionally.  Equation (\ref{eq:FOM}) is  the full-order model (FOM) and the cost associated with its computational solution is prohibitive for a large number of particles (large $n$), and/or chemical reactions with a large number of species (large $n_s$).  To reduce this cost, this equation is solved on a low-rank matrix manifold by approximating  $\Phi$ with a rank-$r\ (r< n_c)$ matrix. To describe the ROM strategy, first, some formal definitions are given below.

\begin{figure}[t]
    \centering
    \includegraphics[width=.5\textwidth]{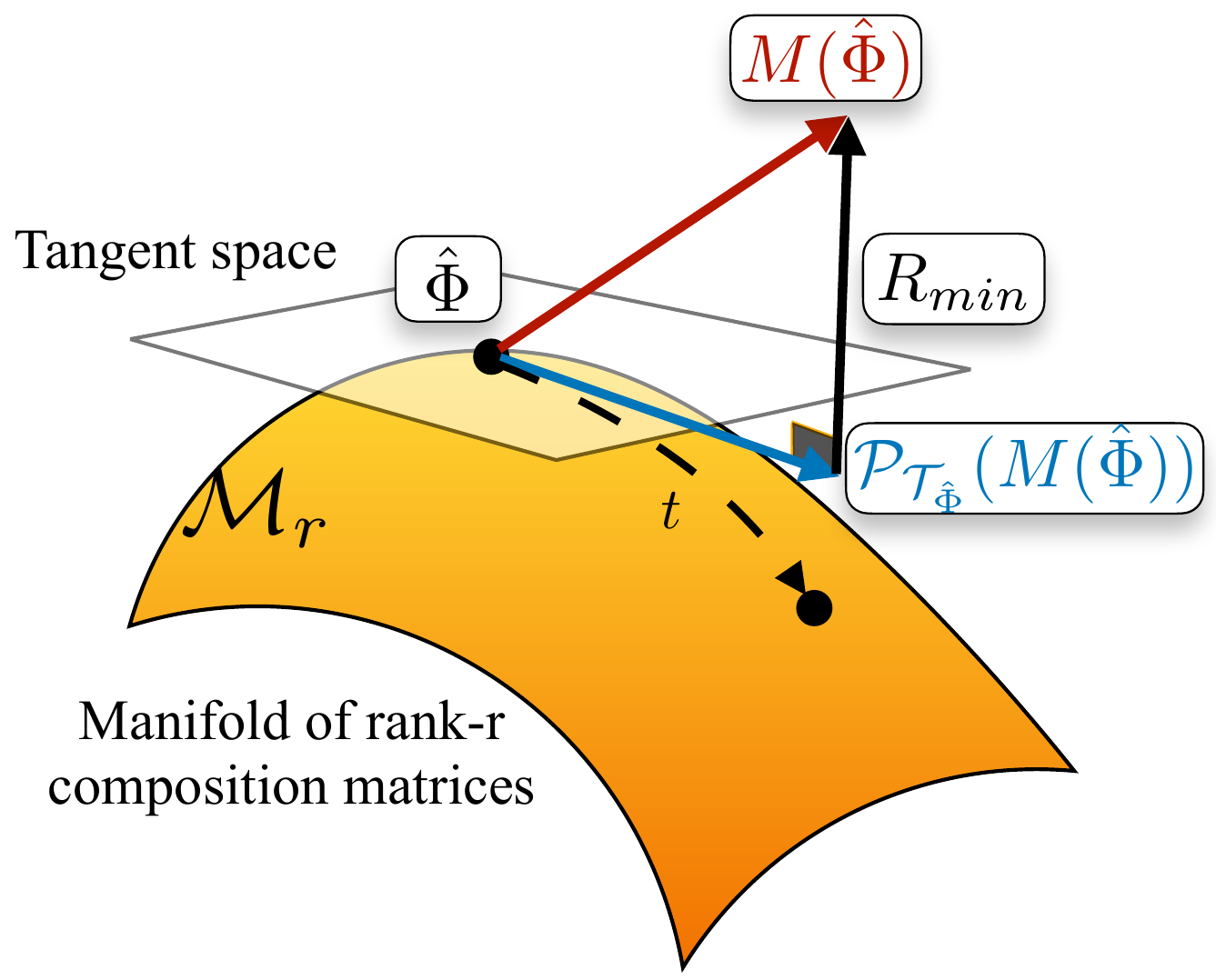}
    \caption{Schematics of the low-rank matrix manifold for on-the-fly reduced-order modeling of composition matrix.}
    \label{fig:Manifold}
\end{figure}

\begin{definition}[Low-rank matrix manifolds]\label{def:Mr}
 The low-rank matrix manifold $\mathcal{M}_r$ is defined as the set
\begin{equation*}
\mathcal{M}_r = \{\hat{\Phi} \in \mathbb{R}^{n \times n_c}: \ \mbox{rank}(\hat{\Phi}) = r \}, 
\end{equation*}
of matrices of fixed rank $r$. Any member of the set $\mathcal{M}_r$ is denoted by a hat symbol $( \hat{ \ \ } )$, e.g., $\hat{\Phi}$.
\end{definition}
The manifold $\mathcal M_r$ is shown in Fig. \ref{fig:Manifold} where any point or vector represents a matrix of size $n \times n_c$. The points that lie on the manifold are rank-$r$ matrices. 
In the presented methodology the rank-$r$ matrix is parameterized as:
\begin{equation}\label{eq:TDB}
   \hat{\Phi}(t) =  U(t) \Sigma (t) Y^\top(t),
\end{equation}
where $U(t) =[u_1(t), u_2(t), \dots, u_r(t)] \in \mathbb{R}^{n \times r}$ is the set of orthonormal vectors, i.e., $u^\top_i(t)u_j(t)= \delta_{ij}$, or $U(t)^\top U(t)  = I_r$, where $I_r \in \mathbb{R}^{r \times r}$ is the identity matrix. Similarly, $Y(t) =[y_1(t), y_2(t), \dots, y_r(t)] \in \mathbb{R}^{n_c \times r}$ is  a set of orthonormal vectors, i.e., $y^\top_i(t)y_j(t)= \delta_{ij}$, or $Y(t)^\top Y(t)  = I_r$. The matrix $Y(t)$ represents a time-dependent subspace in the composition space.  In the above parameterization,  $\Sigma \in \mathbb{R}^{r \times r}$, and  Eq.\ (\ref{eq:TDB}) resembles a rank-\(r\) truncated singular value decomposition (SVD), with the key difference that \(\Sigma\) is not required to be a diagonal matrix as is the case in standard SVD. 

 \begin{definition}[Tangent space]\label{def:Tan_spc}
 The tangent space of manifold $\mathcal{M}_r$ at $\hat{\Phi}$, represented with the decomposition of $\hat{\Phi} = U \Sigma Y^\top$, is  the set of matrices in the form of \cite{KL07}:
\begin{equation*}
\mathcal{T}_{\Phat} \mathcal{M}_r = \{\delta  U  \Sigma  Y^\top +   U \delta \Sigma  Y^\top +    U  \Sigma \delta  Y^\top: \  \delta  U^\top  U =  0 \ \mbox{and} \  \delta  Y^\top  Y =  0 \},
\end{equation*}
where $\delta  U \in \mathbb{R}^{n \times r}$ and $\delta  Y \in \mathbb{R}^{s \times r}$.
\end{definition}

\begin{definition}[Orthogonal projection onto the tangent space]\label{def:ortho_Tan_spc}
 The orthogonal projection of matrix $ W \in \mathbb{R}^{n\times s}$ onto the  tangent space of manifold $\mathcal{M}_r$ at $\hat{ V}$, represented with the decomposition of $\hat{ V} = {U}  \Sigma  Y^\top$, is  given by \cite[Lemma 4.1]{KL07}:
\begin{equation}\label{eq:tan_spc}
    \mathcal{P}_{\mathcal{T}_{\hat{{V}}}} ( W) = {U}{U}^\top  W +  W {Y}{Y}^\top -  {U}{U}^\top  W {Y}{Y}^\top.
\end{equation}
\end{definition}
Replacing $\Phat$ into the FOM results in a residual due to the low-rank approximation error, $\Phi=\Phat +E$, where $E \in \mathbb{R}^{n\times n_c}$ is the low-rank approximation error. As a result,  $\Phat$ does not satisfy Eq.\ (\ref{eq:FOM}) exactly and it will generate a residual as shown below:
\begin{equation}
\frac{d\Phat}{dt} = M (\Phat) + R,
\end{equation}
where $R \in \mathbb{R}^{n\times n_c}$ is the residual matrix. The evaluation equation for the composition matrix in the low-rank form is obtained by finding the optimal $\dot{\Phat} = d\Phat/dt$ that minimizes the norm of the above residual under the constraint that $\Phat$ remain a rank-$r$ matrix, i.e.,  $\Phat \in \mathcal{M}_r$. The constrained minimization is given below:
\begin{equation} \label{eq:VarPrin}
\min_{\dot{\Phat}} \mathcal J(\dot{\Phat}) = \big \| \dot{\Phat} - M(\Phat) \big \|^2_F, \quad \mbox{such that} \quad \Phat \in \mathcal{M}_r,
\end{equation}
where $\mathcal J(\dot{\Phat}) = \big \| R \big \|^2_F$ and $\big \| \ . \ \big \|_F$ is the Frobenius matrix norm:
\begin{equation}
    \| R \|^2_F = \sum_{i=1}^n \sum_{j=1}^{n_c} R^2_{ij}.
\end{equation}
In simple words, the above minimization seeks to find the optimal $\dot{\Phat}$ such that $\Phat$ remains a rank-$r$ matrix. Therefore, Eq.\ (\ref{eq:VarPrin}) is a constrained minimization problem and can be solved using Lagrange multipliers \cite{RNB21}. This problem can alternatively be solved using the Riemannian optimization where Riemannian geometric concepts are utilized to solve the above constrained problem \cite{KL07}. Here, the  Riemannian approach is followed, by which the minimization problem is obtained by the orthogonal projection of $M(\Phat)$ onto the tangent space:
\begin{equation}\label{eq:opt_ev}
\dot{\Phat} = \mathcal{P}_{\mathcal{T}_{\Phat}}(M(\Phat)). 
\end{equation}
It is easy to verify that to ensure $\Phat$ remains on the manifold,  $\dot{\Phat}$ must belong to the tangent space. This can also be understood geometrically: if $\dot{\Phat}$ has any component normal to the tangent plane, it will move $\Phat$ off the manifold. 
Moreover, the optimal $\dot{\Phat}$ that minimizes the residual is obtained by orthogonal projection onto the tangent space. Therefore,
\begin{equation}
    R_{min} = M(\Phat) - \mathcal{P}_{\mathcal{T}_{\Phat}}(M(\Phat)).
\end{equation}
The geometric depiction of the projection onto the tangent space is shown in Fig.\ \ref{fig:Manifold}.
Equation (\ref{eq:opt_ev}) does not immediately lend itself to a cost-effective method for advancing composition transport. The reason is that this equation is formulated versus $\Phat$, whose number of entries is the same as the FOM. However, since $\Phat$ is low-rank, it can be expressed in the SVD-like factorized form as shown in Eq.\ (\ref{eq:TDB}).  From the orthonormality of $U$ basis,  $U^\top U = I_r$: 
\begin{equation}
\frac{d(U^\top U)}{dt} = \dot{U}^\top U + U^\top \dot{U} = 0.
\end{equation}
 Therefore, $\Psi_U = U^\top \dot{U} \in \mathbb{R}^{r \times r}$ is a skew-symmetric matrix, i.e., $\Psi_U^\top=-\Psi_U$. Similarly, $\Psi_Y = Y^\top \dot{Y} \in \mathbb{R}^{r \times r}$ is also a skew-symmetric matrix. As shown in Ref.\ \cite{RNB21}, any skew-symmetric choice for $\Psi_U$ and $\Psi_Y$ results in equivalent low-rank approximations. Here,  the simplest form, i.e., $\Psi_U=\Psi_Y=0$, is used which is also known as the dynamically orthogonal (DO) condition \cite{SL09}. 
Using the factorized form of $\Phat$:
\begin{equation}
 \dot{\Phat} =  \dot{U} \Sigma Y^\top + U \dot{\Sigma} Y^\top + U \Sigma \dot{Y}^\top.
\end{equation}
Using this into Eq.\ (\ref{eq:opt_ev}) yields:
\begin{equation}\label{eq:tan_space_proj}
\dot{U} \Sigma Y^\top + U \dot{\Sigma} Y^\top + U \Sigma \dot{Y}^\top = {U}{U}^\top  M +  M {Y}{Y}^\top -  {U}{U}^\top  M {Y}{Y}^\top. 
\end{equation}
where $M\equiv M(U\Sigma Y^\top)$ is used for simplicity. 
Multiplying Eq.\ (\ref{eq:tan_space_proj})  from left by $U^\top$ and from right by $Y$ gives:
\begin{equation}\label{eq:sigma_dot}
\dot{\Sigma}  = {U}^\top  M Y+  U^\top M {Y} -  {U}^\top  M {Y} = U^\top M Y,
\end{equation}
where  the orthonormality of $U$ and $Y$ are imposed, i.e., $U^\top  U = I$ and $Y^\top  Y = I$.  Multiplying Eq.\ (\ref{eq:tan_space_proj}) from left by $U^\top$ results in:
\begin{equation}
     \dot{\Sigma} Y^\top + \Sigma \dot{Y}^\top = {U}^\top  M +  U^\top M {Y}{Y}^\top -  {U}^\top  M {Y}{Y}^\top. 
\end{equation} 
Replacing  $\dot{\Sigma}$ from Eq.\ (\ref{eq:sigma_dot}) and simplifying the above equation results in:
\begin{equation}
   \Sigma \dot{Y}^\top = {U}^\top  M  -  {U}^\top  M {Y}{Y}^\top = ({U}^\top  M)(I_{n_c} -Y Y^\top). 
\end{equation}
where $I_{n_c}$ is the identity matrix of size $n_c \times n_c$. Transposing  the above equation and multiplying both sides by $ \Sigma^{-T}$ gives:
\begin{equation}
   \dot{Y} = (I_{n_c} -Y Y^\top) (M^\top U) \Sigma^{-T}.  
\end{equation}
Similarly, multiplying Eq.\ (\ref{eq:tan_space_proj}) from right by $Y$ yields:
\begin{equation}
\dot{U} \Sigma + U \dot{\Sigma}  = {U}{U}^\top  M Y +  M {Y} -  {U}{U}^\top  M {Y}.
\end{equation}
In this equation,   $\dot{\Sigma}$ can be replaced from Eq.\ (\ref{eq:sigma_dot}).  Therefore:  
\begin{equation}
\dot{U}    =  (M Y -  {U}{U}^\top  M {Y})\Sigma^{-1} = (I_n-UU^\top)MY\Sigma^{-1},
\end{equation}
where $I_n$ is the identity matrix of size $n \times n$. The final  evolution equations for $U$, $\Sigma$ and $Y$ are:
\begin{subequations}
\label{eq:evol_USY_general}
\begin{align}
\dot{U}    &= (I_n-UU^\top)M(U\Sigma Y^\top)Y\Sigma^{-1}, \label{eq:Uevol}\\
\dot{Y}  &= (I_{n_c} -Y Y^\top) M^\top(U\Sigma Y^\top) U \Sigma^{-T}, \label{eq:Yevol}\\
\dot{\Sigma}  &= U^\top M(U\Sigma Y^\top) Y. \label{eq:Sevol}
\end{align}
\end{subequations}
These equations constitute the evolution equation of $\Phat$ in the low-rank form and constitute the framework for  TDB-ROM.  

\section{Computational Complexity}

The computational complexity is considered in terms of memory and floating-point operations (flops). The TDB-ROM evolution equation enables the solution of the composition transport equation in a compressed form by reducing its degrees of freedom. The total number of entries in the factorized form of this equation is \(rn + r^2 + rn_c \approx \mathcal{O}(rn)\), since \(r \ll n\) and \(n_c \ll n\). This is to be compared with that in FOM, which requires tracking of the matrix \(\Phi\) with \(n_c n\) entries. Therefore, TDB-ROM results in memory savings by an approximate factor of \(n_c n/rn = n_c/r\). 

Memory compression is demonstrated by considering Eqs. (\ref{eq:Uevol})-(\ref{eq:Sevol}) that involve the computations of
$ M \equiv M(U\Sigma Y^\top), $ a matrix of size $n \times n_c$. If this matrix is explicitly formed, the memory compression vanishes as this requires storing $n_c n$ entries, the same as that in FOM. However, it is possible to solve the ROM  without storing all entries of the matrix at once. The key observation is that only the  projections of $M$ onto the $U$ and $Y$ subspaces are required. Specifically: $ M_Y = MY \in \mathbb{R}^{n \times r} \quad \text{and} \quad M_U = U^\top M \in \mathbb{R}^{r \times n_c} $. The memory storage required for these arrays is $\mathcal{O}(rn)$ or smaller. The following pseudocode shows how $M_Y$ and $M_U$ can be computed without explicitly forming or storing the matrix $M$ in memory by summing over the appropriate indices:
\begin{align*}
    &M_{U} = \mathbf{0} \\
    &\texttt{for } i = 1 : n \\
    &\quad M_{Y}(i,:) = M(i,:) * Y \\
    &\quad M_{U} = M_{U} + U(i,:)^\top * M(i,:) \\
    &\texttt{end}
\end{align*}
Here MATLAB indexing syntax is used in which $M(i,:)$ denotes the $i$th row of matrix $M$, and $*$ denotes matrix-to-matrix multiplications. The flops cost of solving TDB-ROM scales with $\mathcal{O}(r n_c n)$. For demonstration, consider the cost of computing both terms in $M(\Phat) = -L(\Phat) + S(\Phat)$. The first term $L(\Phat) = \Omega \left( \Phat - \la \Phat\ra_{L} \right)$ is  linear  and can be expressed as: $L(\Phat) = D \Phat,$
where $D = \Omega (I_n -A)  \in \mathbb{R}^{n \times n}$ and $A  \in \mathbb{R}^{n \times n}$ is a sparse matrix representing the average operation ($ \la \cdot \ra_L $):
$$ D \Phat = D U \Sigma Y^\top =  \Omega (U - AU)\Sigma Y^\top = \Omega (U -   \la U \ra_{L} )\Sigma Y^\top $$
Therefore, computing $D \Phat$ requires computation of $ \la U \ra_{L}$, which requires $\mathcal{O}(rn)$ flops. The term $\Sigma Y^\top $ should \emph{not} be multiplied to $DU$ as $\Sigma Y^\top $ is simplified after exerting $ D \Phat$ into Eqs.\ (\ref{eq:Uevol})-(\ref{eq:Sevol}):

\begin{align*}
    (I_n-UU^\top)L(\Phat)Y\Sigma^{-1} &= (I_n-UU^\top)D \Phat Y\Sigma^{-1}\\
                                             &= (I_n-UU^\top)D U \Sigma Y^\top Y\Sigma^{-1}\\
                                             & = (I_n-UU^\top)D U\\
                                             & = DU -UU^\top D U.
\end{align*}
Computing $UU^\top D U$ scales with $\mathcal{O}(r^2 n)$.  Similarly, 
\begin{align*}
    (I_{n_c} -Y Y^\top) L(\Phat)^\top U \Sigma^{-T} &= (I_{n_c} -Y Y^\top) (DU \Sigma Y^\top)^\top U \Sigma^{-T}\\
                                             &= (I_{n_c} -Y Y^\top)Y\Sigma^\top U^\top D^\top  U \Sigma^{-T}\\
                                             & = 0,
\end{align*} 
since $(I_{n_c} -Y Y^\top)Y= 0$. In other words, the linear term does not contribute to the evolution of $Y$. If the computation of the chemical source term for each particle scales linearly with the number of scalars, the computation of $S(\Phat) = S(U \Sigma Y^\top)$ scales with $\mathcal{O}(n_c n)$. Calculating the contribution of the source term to the $M_U$ and $M_Y$ scales with $\mathcal{O}(r n_c n)$. As a result, the computation of the chemical source term and its projection coefficients onto the
$U$ and $Y$ subspaces become the dominant cost, especially when dealing with a large number of scalar variables. 

In summary, the TDB methodology developed here provides substantial computational memory savings in conducting LES-FDF. However, the extent of reduction in flops is not significant.  This is simply  because the matrix \( S(U \Sigma Y^\top) \) remains full-rank, despite \( U \Sigma Y^\top \) being low-rank.  Interestingly, the matrix \( S(U \Sigma Y^\top) \) itself is highly amenable to accurate low-rank approximations. To reduce the flops cost an interpolatory low-rank approximation of \( S(U \Sigma Y^\top) \) is possible.  This has been   recently developed in Refs.\ \cite{NB23,DPNFB23}.  Implementation of these approximations in LES-FDF-TDB is an ambitious task, but would be very valuable.

\section{Simulations}

\begin{figure}
    \centering
    \includegraphics[width=0.49\textwidth]{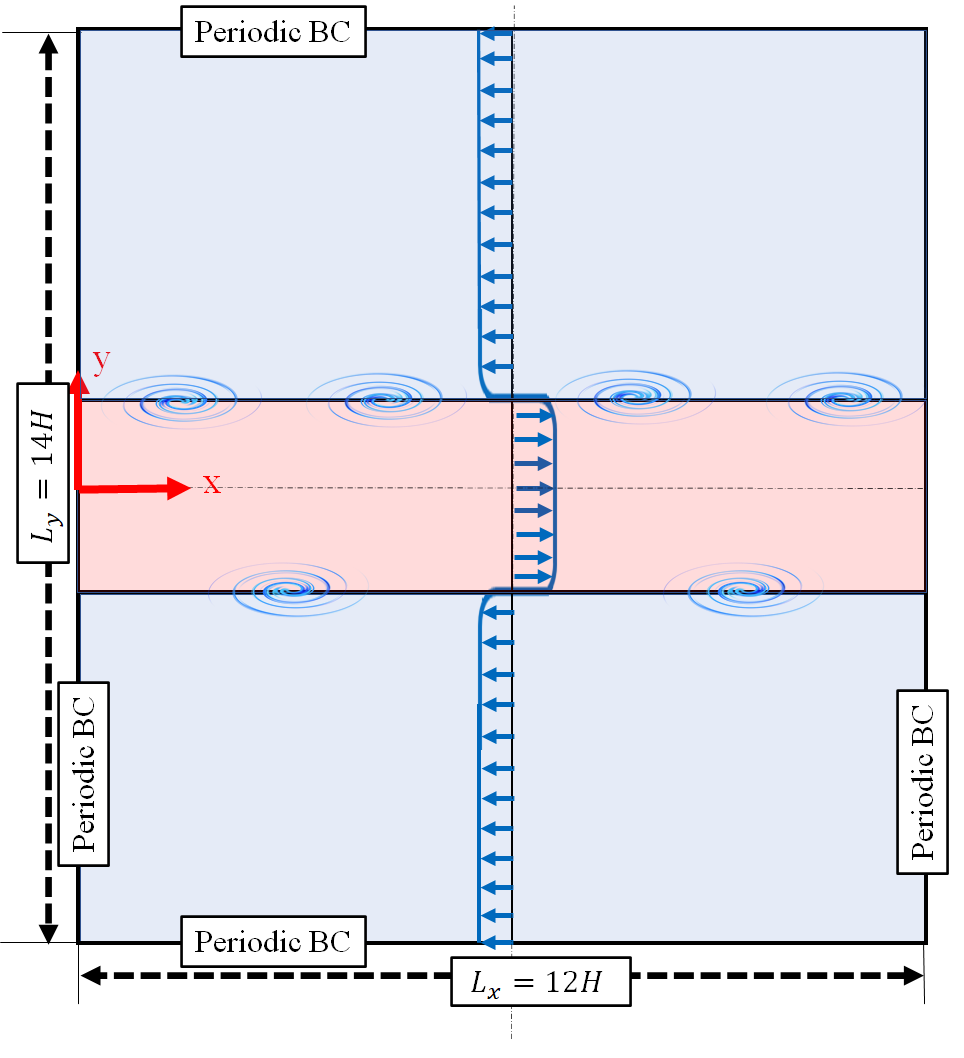}
    \caption{Schematics of the temporally developing turbulent jet flame.}
    \label{fig:FLOW}
\end{figure}

For demonstration, LES is conducted of the canonical configuration of a temporally developing planar CO/H\textsubscript{2} jet flame. The flame is rich with strong flame-turbulence interactions resulting in local extinction followed by re-ignition. This flow configuration has been the subject of previous detailed DNS \cite{Hawkes2007Scalar} and several subsequent modeling and simulations \cite{YANG2013Large, Punati2011Evaluation, Vo2018MMC, Yang2017Sensitivity, Sen2010Large, Aitzhan2022CTM,AITZHAN2025}. These studies indicate that LES-FDF provides a very accurate means of predicting the compositional structure of this flame.  Moreover, the statistics pertaining to non-equilibrium effects in this flame can be predicted via two-dimensional simulations with an excellent accuracy.  Therefore, the two-dimensional configuration as depicted in  Fig.\ \ref{fig:FLOW} is considered here.  The jet consists of a central fuel stream of width $H=0.72$mm surrounded by counter-flowing oxidizer streams. The fuel stream is comprised of 50$\%$ of CO, 10$\%$ H\textsubscript{2} and 40$\%$ N\textsubscript{2} by volume, while oxidizer streams contain 75$\%$ N\textsubscript{2} and 25$\%$ O\textsubscript{2}. The initial temperature of both streams is 500K and thermodynamic pressure is set to 1 atm. The velocity difference between the two streams is $U = 145$m/s. The fuel stream velocity and the oxidizer stream velocity are $U/2$ and $-U/2$, respectively. The initial conditions for the velocity components and mixture fraction are taken directly from center-plane DNS in Ref.\ \cite{Hawkes2007Scalar}, and then the spatial fields of species and temperature are reconstructed from the flamelet table generated with $\chi = 0.75 \chi_{\text{crit}}$, where $\chi$ and $\chi_{\text{crit}}$ are scalar dissipation rate and its critical value, respectively. All of the non-equilibrium effects in this flame can be captured by two-dimensional simulations, and are conducted here. 
 The boundary conditions are periodic in stream-wise ($x$) and cross-stream wise ($y$) directions. The Reynolds number, based on $U$ and $H$ is $Re=2510$. The sound speeds in the fuel and the oxidizer streams are denoted by $C_1$ and $C_2$, respectively and the Mach number $Ma = U / \left(C_1 + C_2 \right) \approx 0.16$ is small enough to justify a low Mach number approximation.   The combustion chemistry is modeled via the skeletal kinetics, containing 11 species with 21 reaction steps \cite{Hawkes2007Scalar}.

 All simulations are conducted via a hybrid finite-difference-MC method as detailed in Refs.\ \cite{MJG06,AJSG11,NNGLP17}. 
 The parallel MC-FDF methodology \cite{APSG12,AGSG11,AJSG11} in this solver allows efficient simulation of the high Reynolds and low Mach number flow.  The size of the computational domain is $L_x \times L_y = 12H \times 14H $. The time is normalized by $t_j = H/U$. The domain is discretized into equally spaced structured fixed grids of size $N_x \times N_y = 144 \times 168 $. The resolution, as selected, is the largest that was conveniently available, and kept the SGS energy within the allowable $15\% \sim 20\%$ of the total energy. The sizes of the ensemble domain, the filter width and the grid sizes are   $\Delta_E = 0.5 \Delta_G = \Delta x = \Delta y = L_x / N_x$. The number of MC particles per grid point is set to 64; so over 1.5 million MC particles portray the FDF at all times.  

The simulated results are analyzed both instantaneously and statistically. In the former, the instantaneous contours (snap-shots) of the reactive scalar fields are considered. In the
latter, the ``Reynolds-averaged'' statistics are constructed. With the assumption of a temporally developing layer, the flow is homogeneous in $x-$ direction. Therefore, all of the Reynolds averaged values, denoted by an overline, are temporally evolving and determined by ensemble averaging over the streamwise direction. The resolved stresses are denoted by $R\lp a,b \rp=\ov{\la a \ra_{L} \la b \ra_{L}} - \lp \ov{\la a \ra_{L}} \rp \lp \ov{\la b \ra_{L}}\rp $, and the total stresses are denoted by $ r\lp a,b \rp= \ov{\lp ab \rp} - \ov{a} \ov{b}$. In LES with the assumption of a {\sl generic} filter, \textit{i.e.}\ $\ov{\la Q \ra_{L}} = \ov{Q}$, the total stresses are approximated by $r_{\text{LES}} \lp a,b \rp =R\lp a,b \rp + \ov{\tau \lp a,b \rp}$ \cite{Germano96,VNPM13}, where $\tau \lp a, b \rp$ is the subgrid variance. The root mean square (RMS) values are square roots of these stresses. 
To analyze the compositional flame structure, the ``mixture fraction'' field $Z (\bx ,t)$ is also constructed. Bilger's formulation \cite{Bilger1976Structure, Peters00} is employed for this purpose.

\begin{figure}
    \centering
     \begin{subfigure}[b]{0.49\textwidth}
        \centering
        \includegraphics[width=\textwidth]{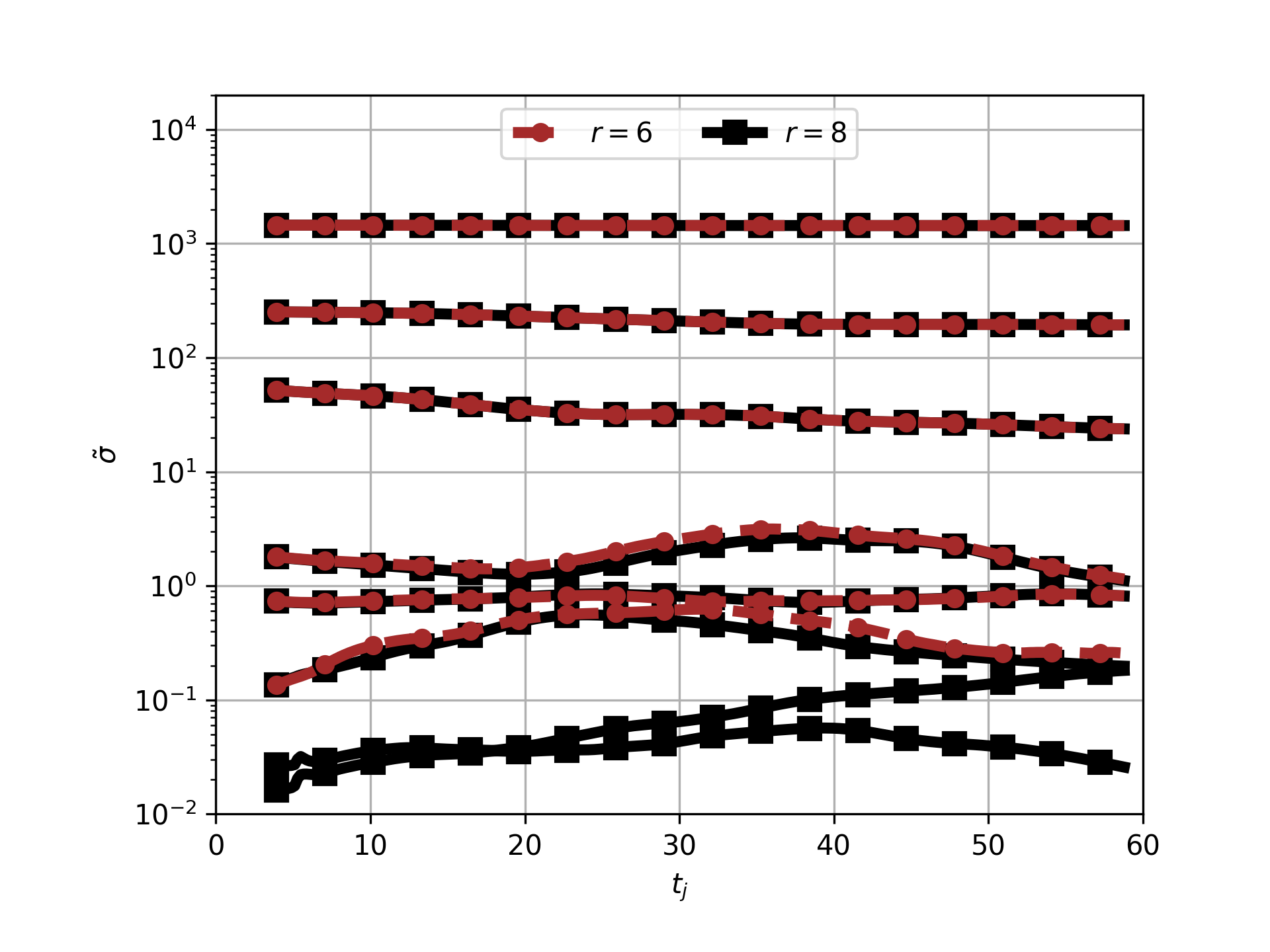} 
        \caption{}
        \label{fig:SIGMA}
    \end{subfigure}
    \hfill
    \begin{subfigure}[b]{0.49\textwidth}
        \centering
        \includegraphics[width=\textwidth]{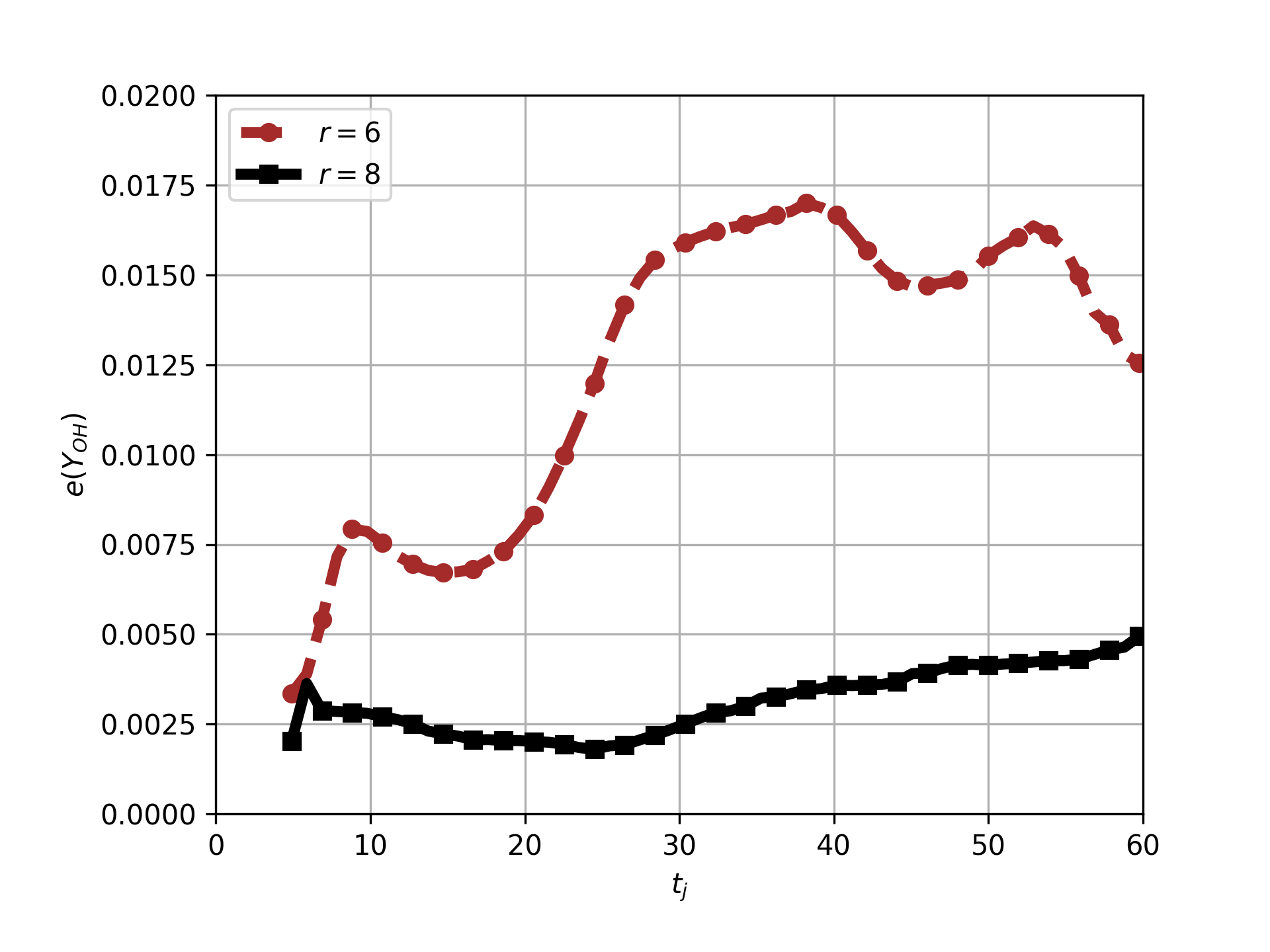}
        \caption{}
        \label{fig:ERRORS}
    \end{subfigure}

    \caption{(a) Temporal evolution of singular values $\Sigma$; (b) temporal evolution of low-rank approximation error in $Y_{OH}$ ($e(Y_{OH}) = \left\lVert Y_{OH}^{FOM} - Y_{OH}^{TDB} \right\rVert_{F}$) with reduction orders of $r=6$ and $r=8$. }
    \label{fig:SIGMA_ERRORS}
\end{figure}

\begin{figure}
    \centering
    \centering
    \begin{subfigure}[b]{0.49\textwidth}
        \centering
        \includegraphics[width=\textwidth]{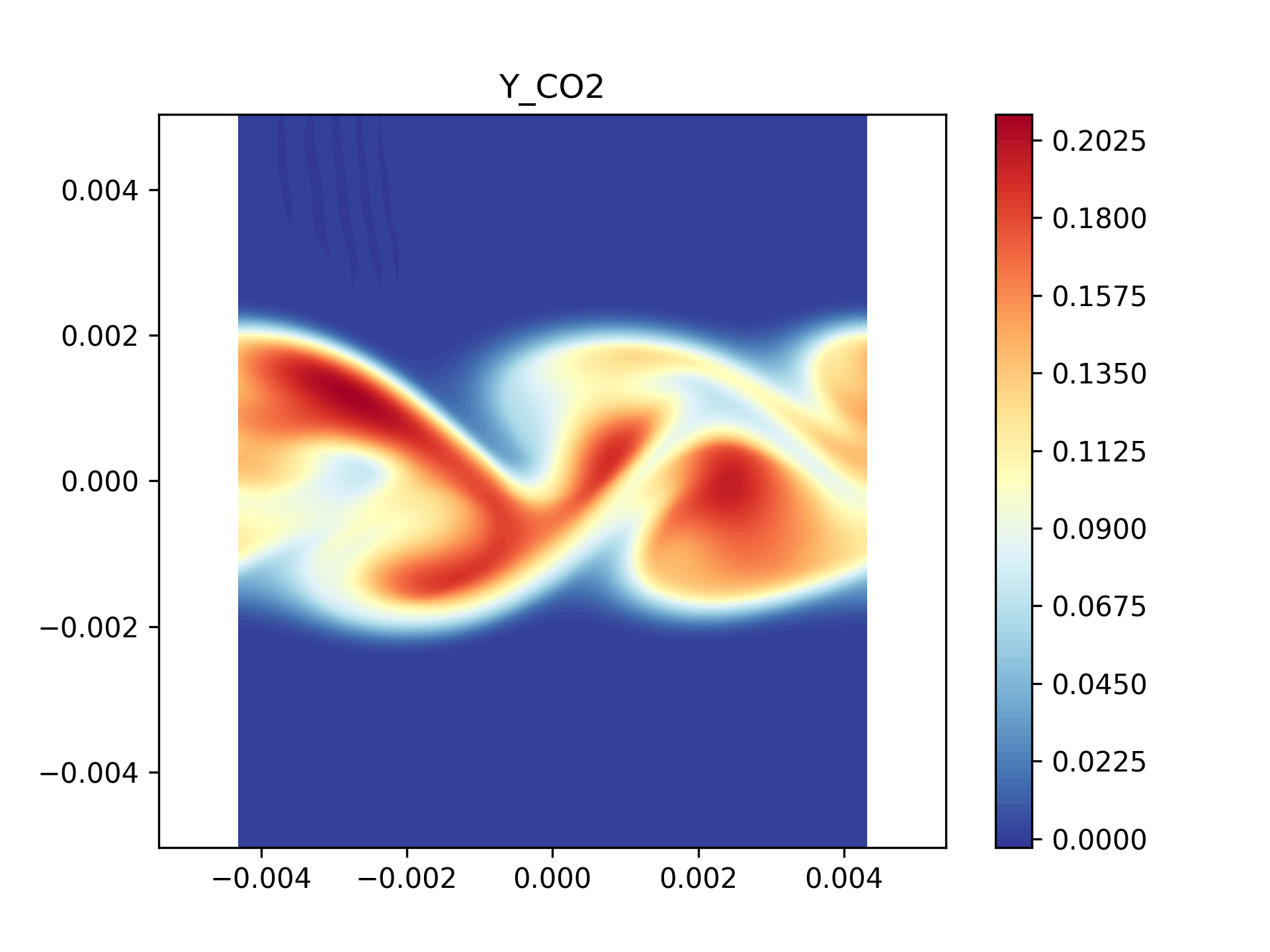}
        \caption{FOM}
    \end{subfigure}

    \centering
    \begin{subfigure}[b]{0.49\textwidth}
        \centering
        \includegraphics[width=\textwidth]{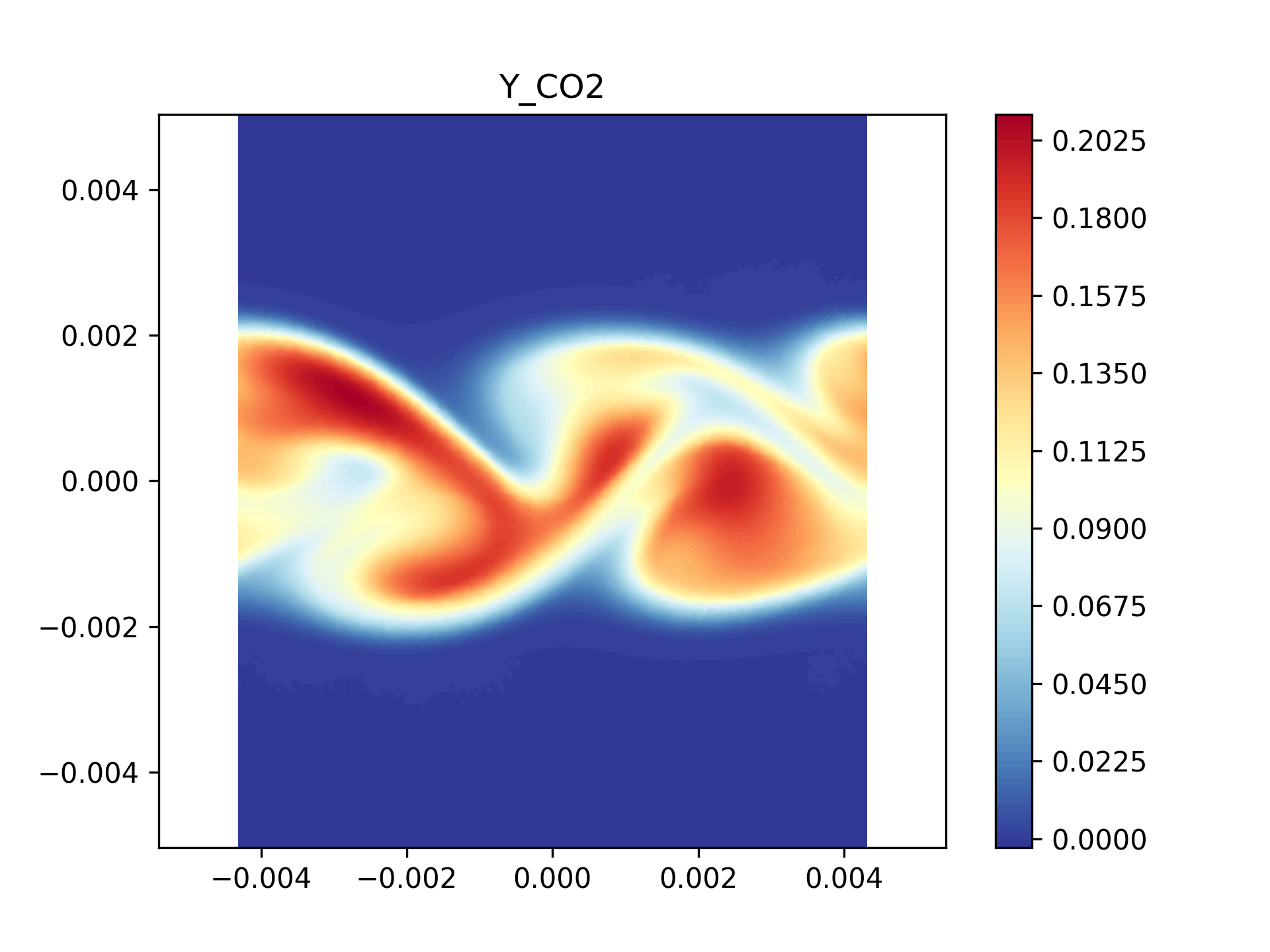}
        \caption{$r=6$}
    \end{subfigure}
    \begin{subfigure}[b]{0.49\textwidth}
        \centering
        \includegraphics[width=\textwidth]{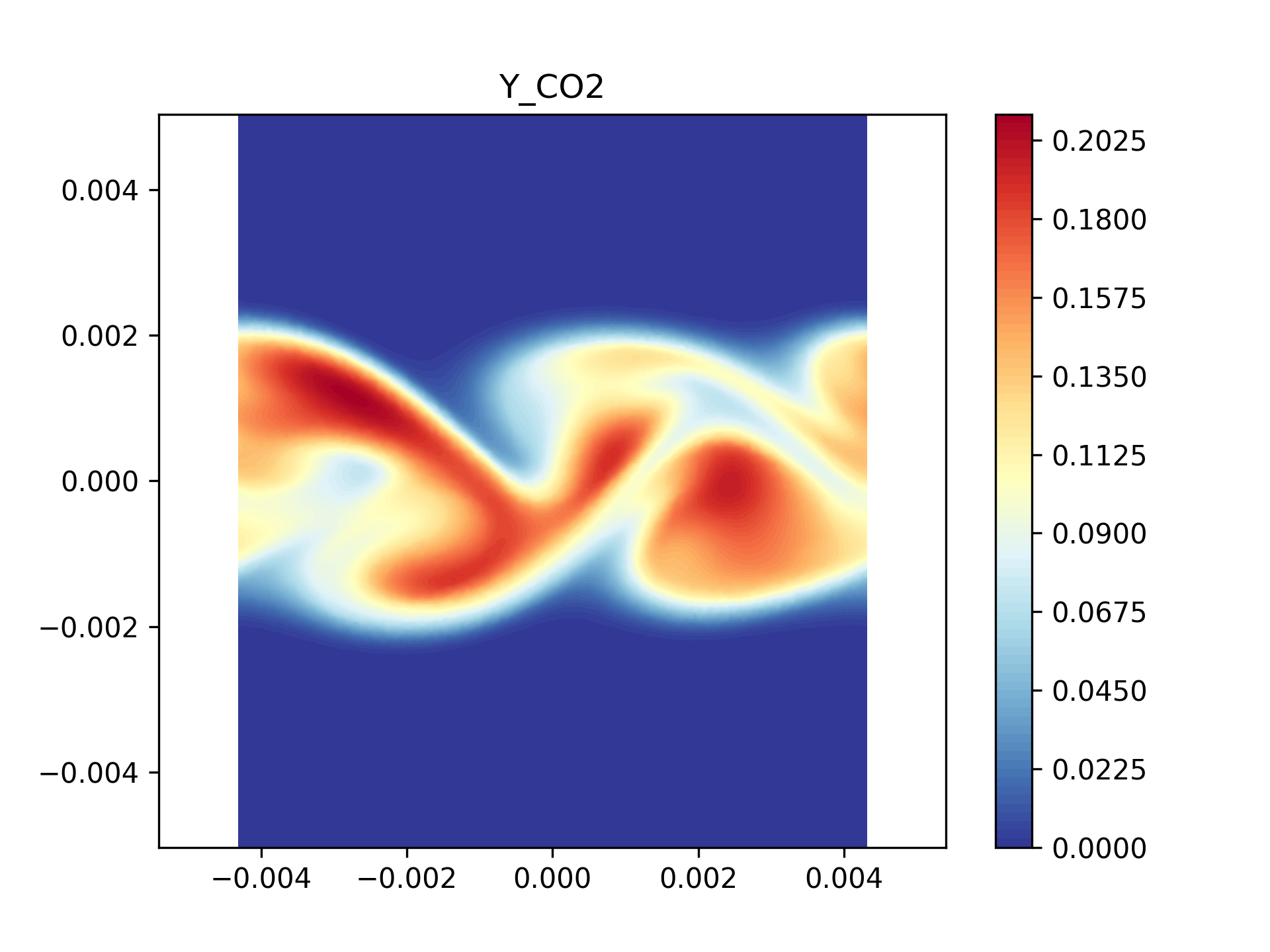}
        \caption{$r=8$}
    \end{subfigure}
    \caption{Instantaneous $CO_2$ field at $t=30t_j$.}
    \label{fig:CO2Field}
\end{figure}

\begin{figure}
    \centering
    \begin{subfigure}[b]{0.49\textwidth}
        \centering
        \includegraphics[width=\textwidth]{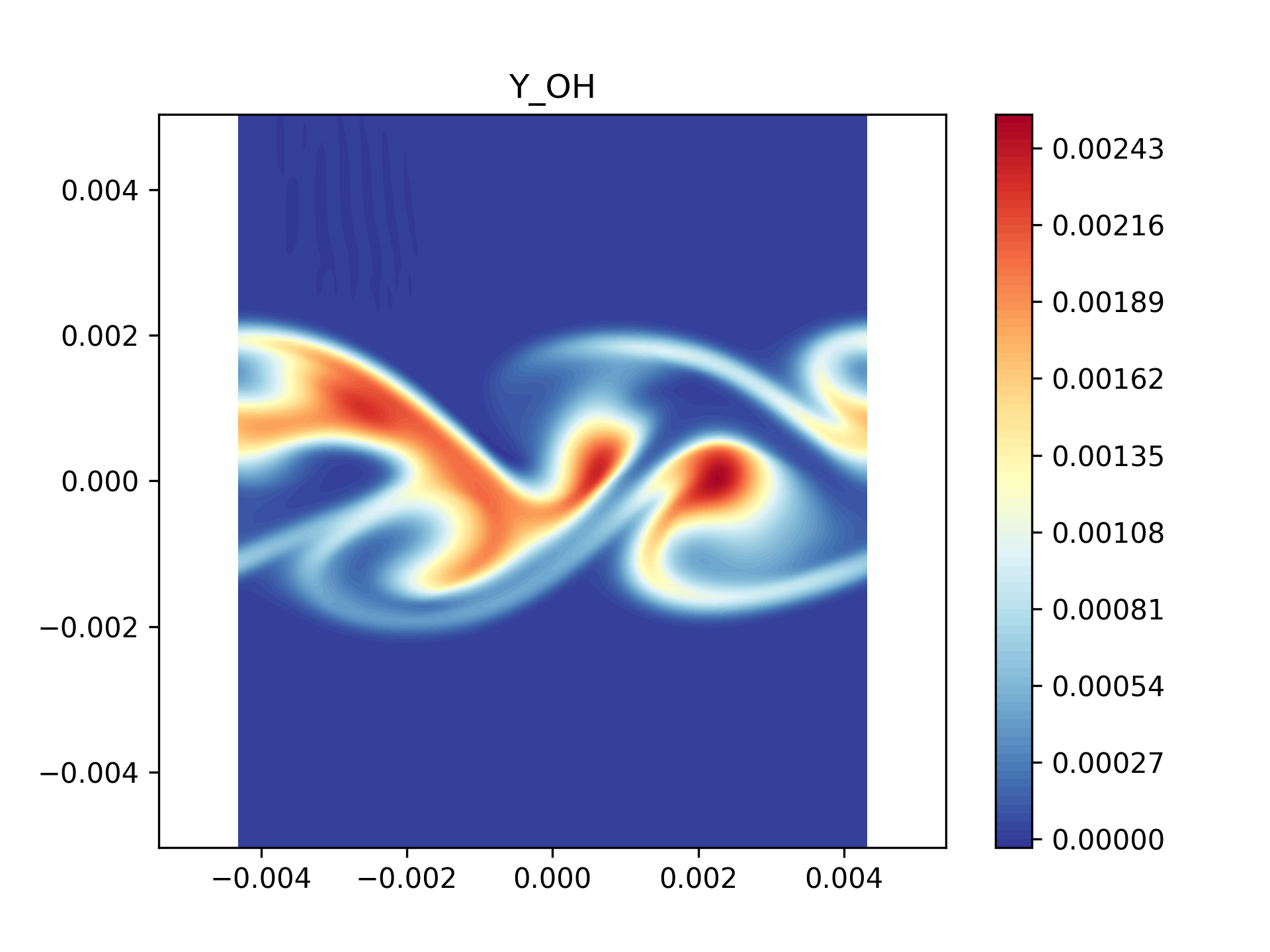}
        \caption{FOM}
    \end{subfigure}

    \begin{subfigure}[b]{0.49\textwidth}
        \centering
        \includegraphics[width=\textwidth]{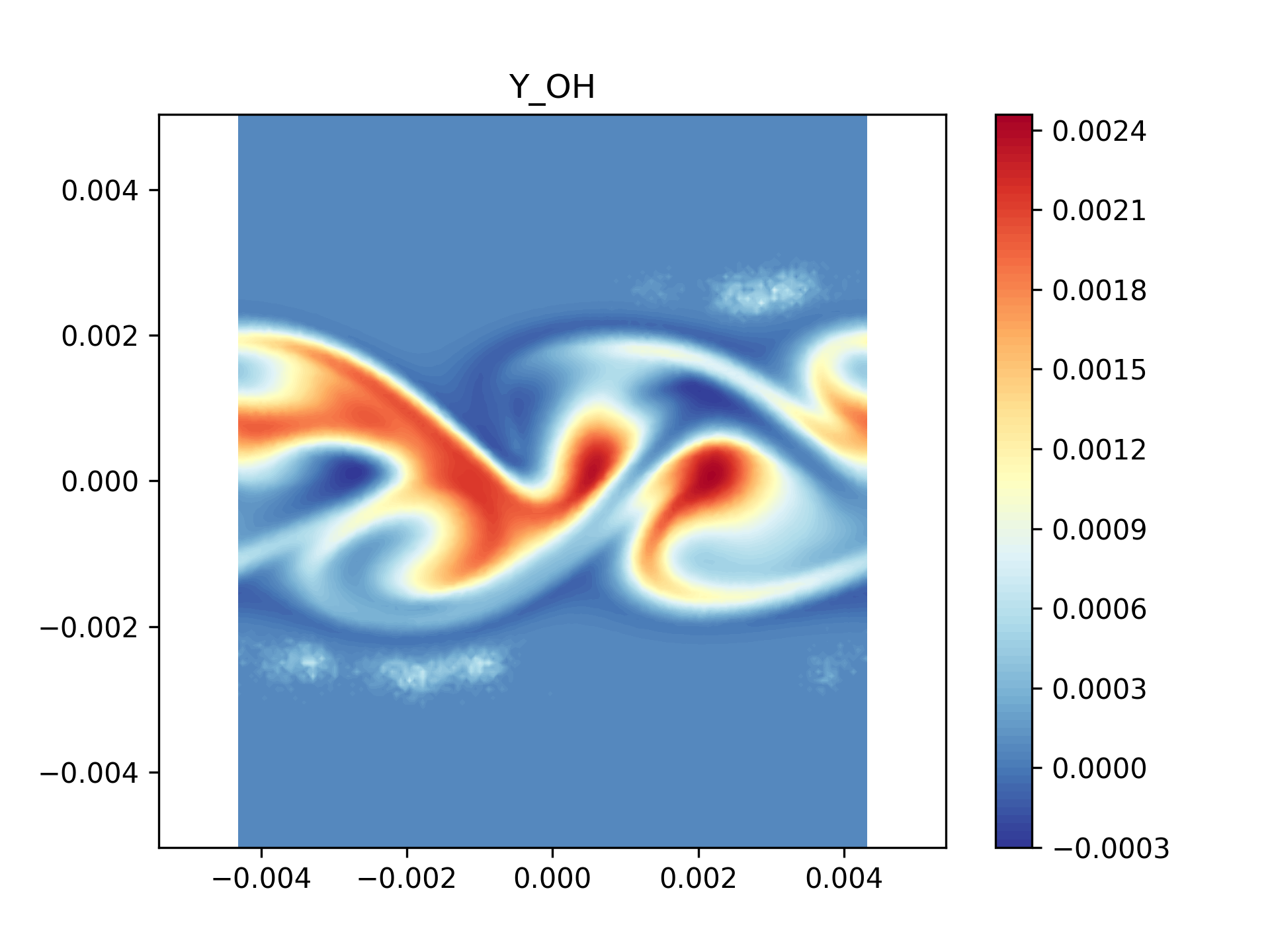}
        \caption{$r=6$}
    \end{subfigure}
    \begin{subfigure}[b]{0.49\textwidth}
        \centering
        \includegraphics[width=\textwidth]{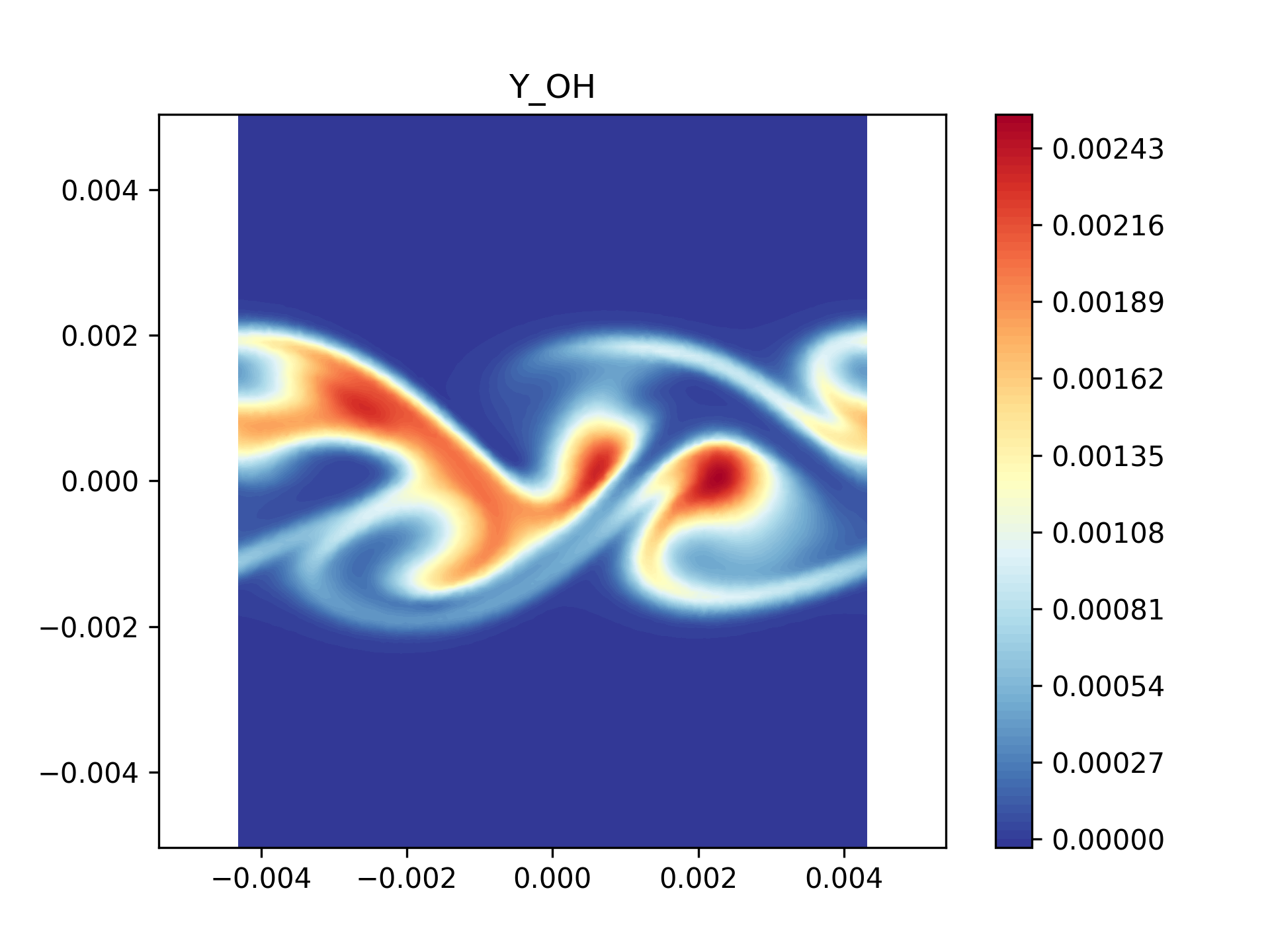}
        \caption{$r=8$}
    \end{subfigure}
    \caption{Instantaneous OH field at $t=30t_j$.}
    \label{fig:OHField}
\end{figure}

\begin{figure}
    \centering
     \begin{subfigure}[b]{0.49\textwidth}
        \centering
        \includegraphics[width=\textwidth]{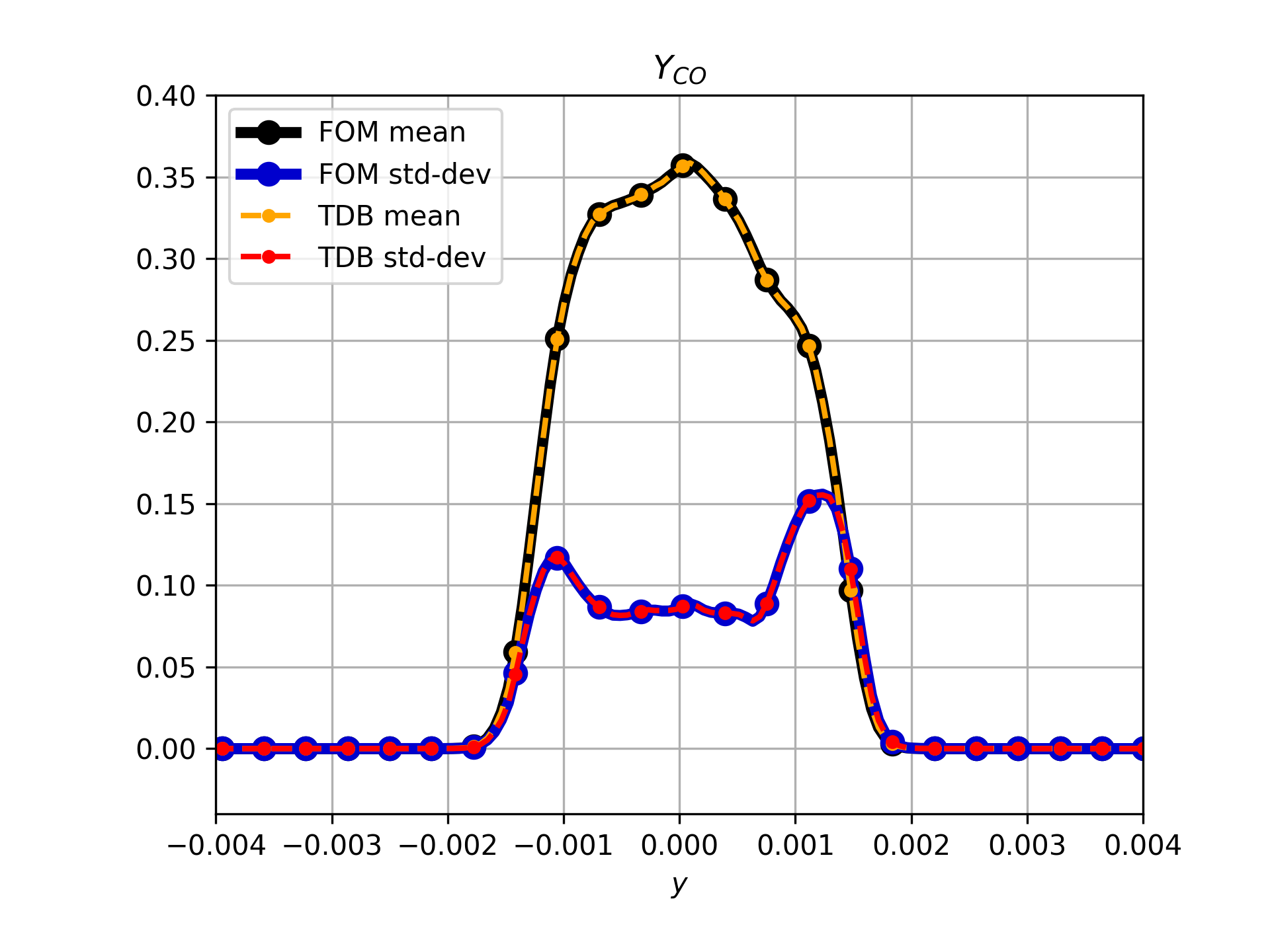}
        \caption{$t=20t_j$}
    \end{subfigure}
     \begin{subfigure}[b]{0.49\textwidth}
        \centering
        \includegraphics[width=\textwidth]{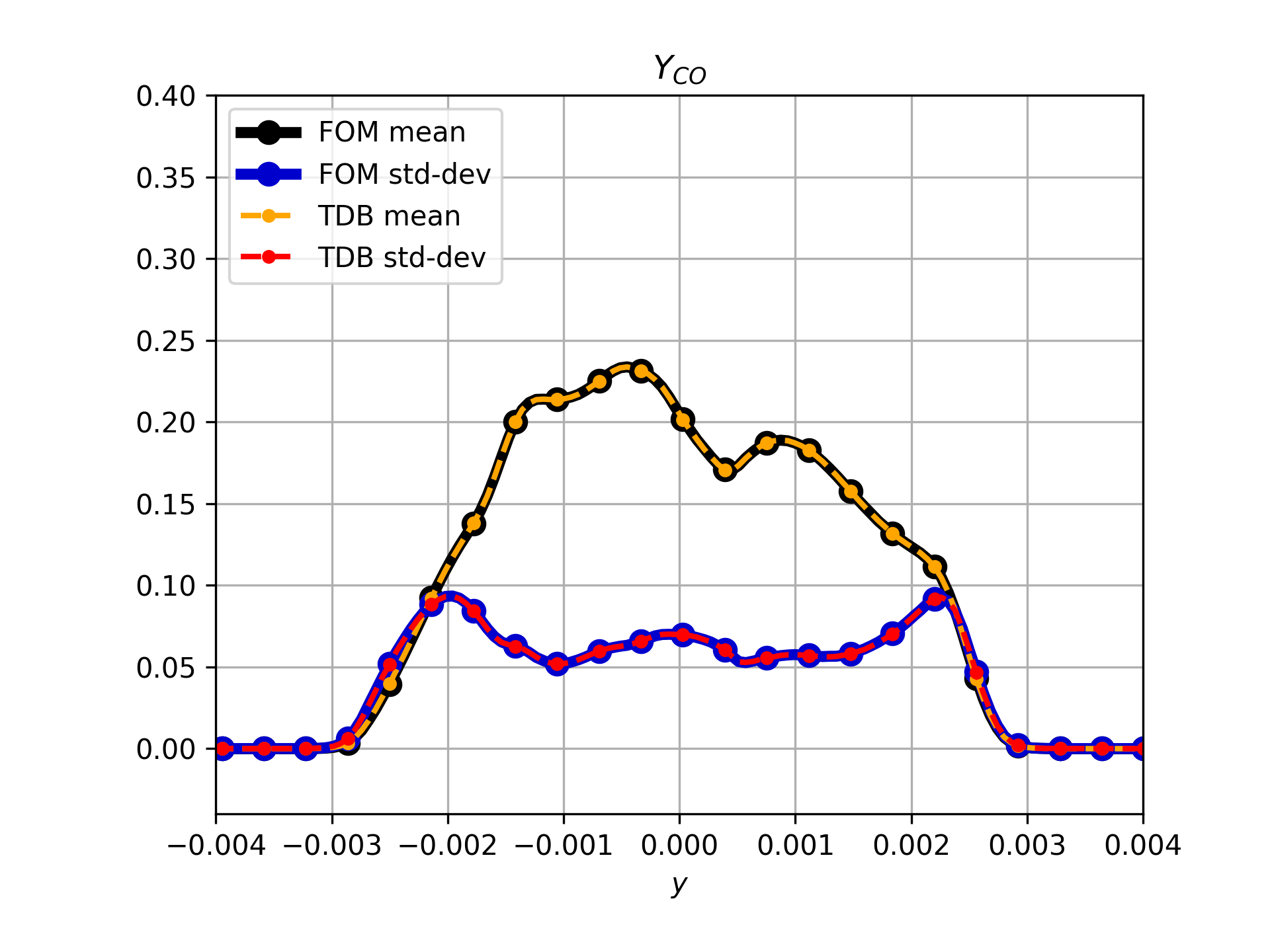}
        \caption{$t=40t_j$}
    \end{subfigure}
    \caption{Reynolds averages of $CO$ (TDB with $r=6$).}
    \label{fig:RavCO}
\end{figure}

\begin{figure}
    \centering
     \begin{subfigure}[b]{0.49\textwidth}
        \centering
        \includegraphics[width=\textwidth]{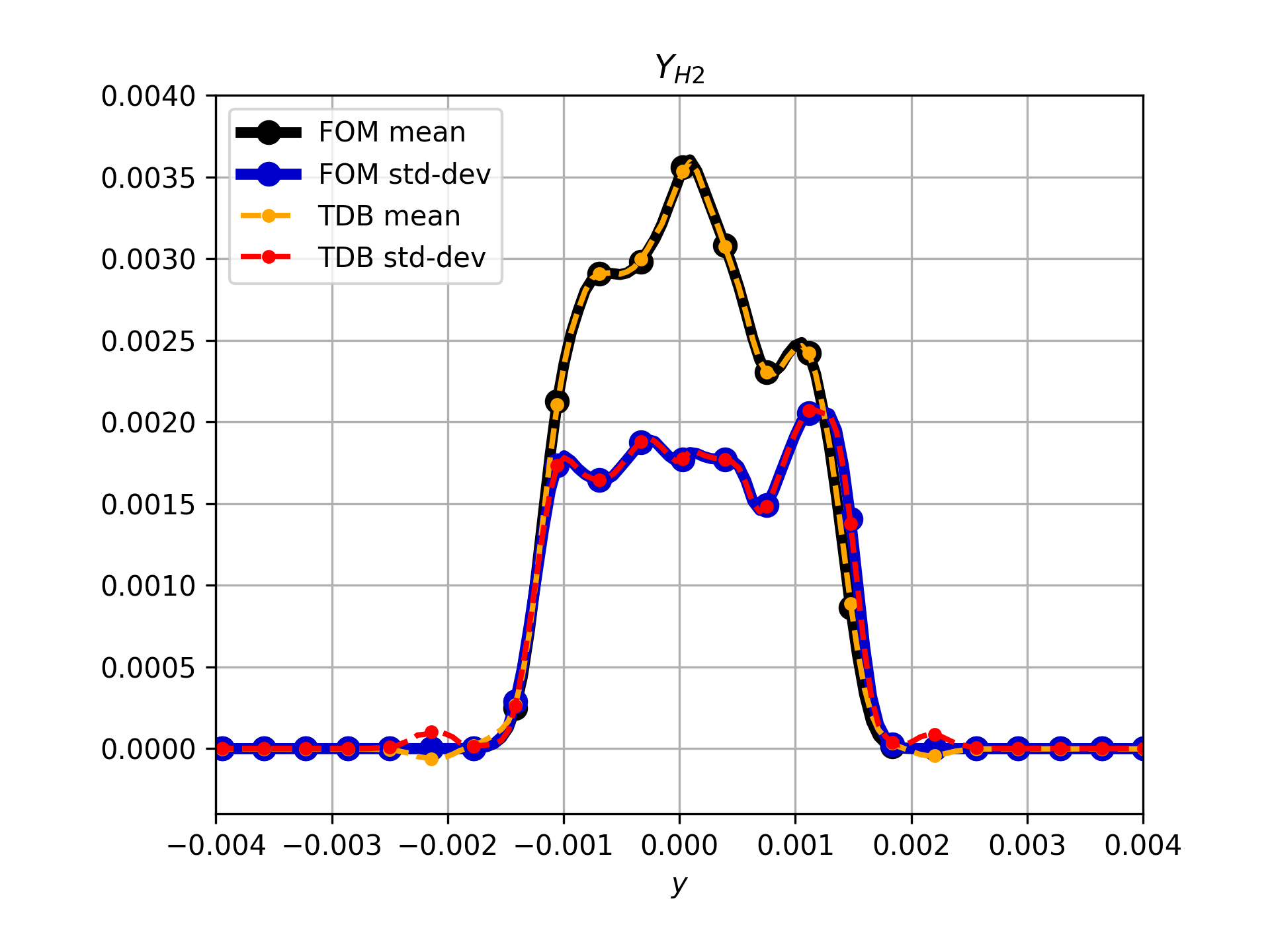}
        \caption{$t=20t_j$}
    \end{subfigure}
     \begin{subfigure}[b]{0.49\textwidth}
        \centering
        \includegraphics[width=\textwidth]{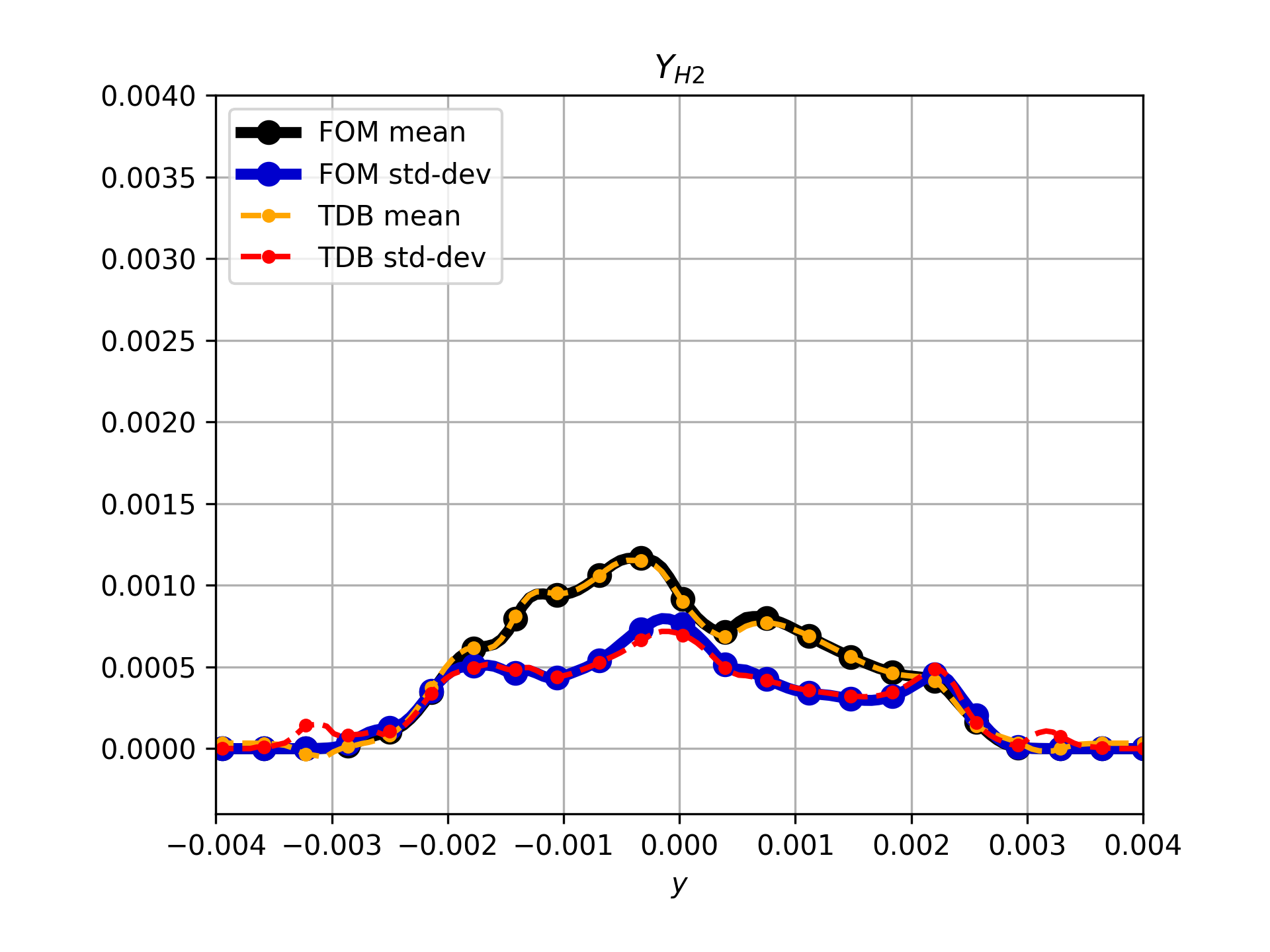}
        \caption{$t=40t_j$}
    \end{subfigure}
    \caption{Reynolds averages of $H_2$ (TDB with $r=6$).}
    \label{fig:RavH2}
\end{figure}

\begin{figure}
    \centering
     \begin{subfigure}[b]{0.49\textwidth}
        \centering
        \includegraphics[width=\textwidth]{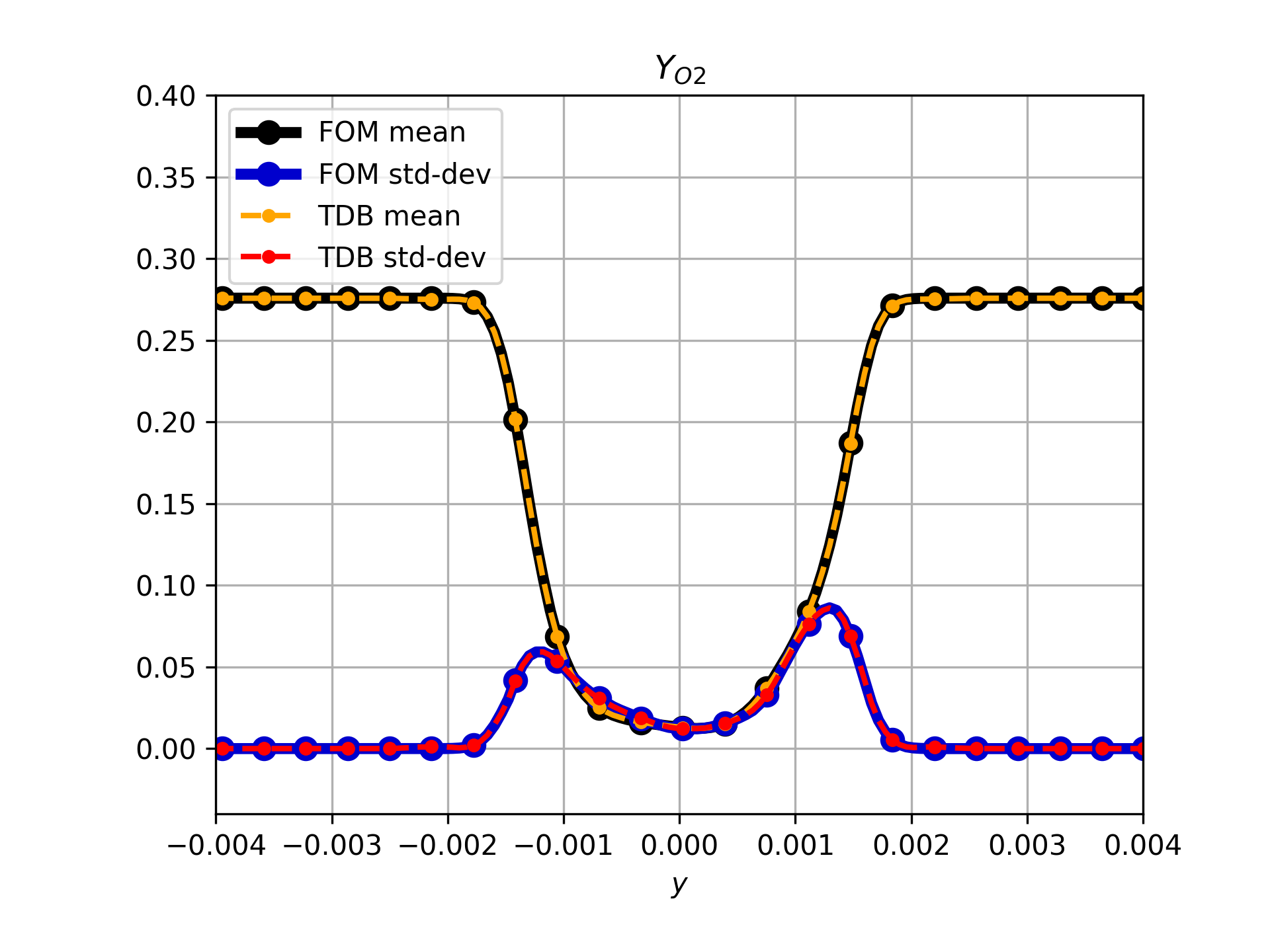}
        \caption{$t=20t_j$}
    \end{subfigure}
     \begin{subfigure}[b]{0.49\textwidth}
        \centering
        \includegraphics[width=\textwidth]{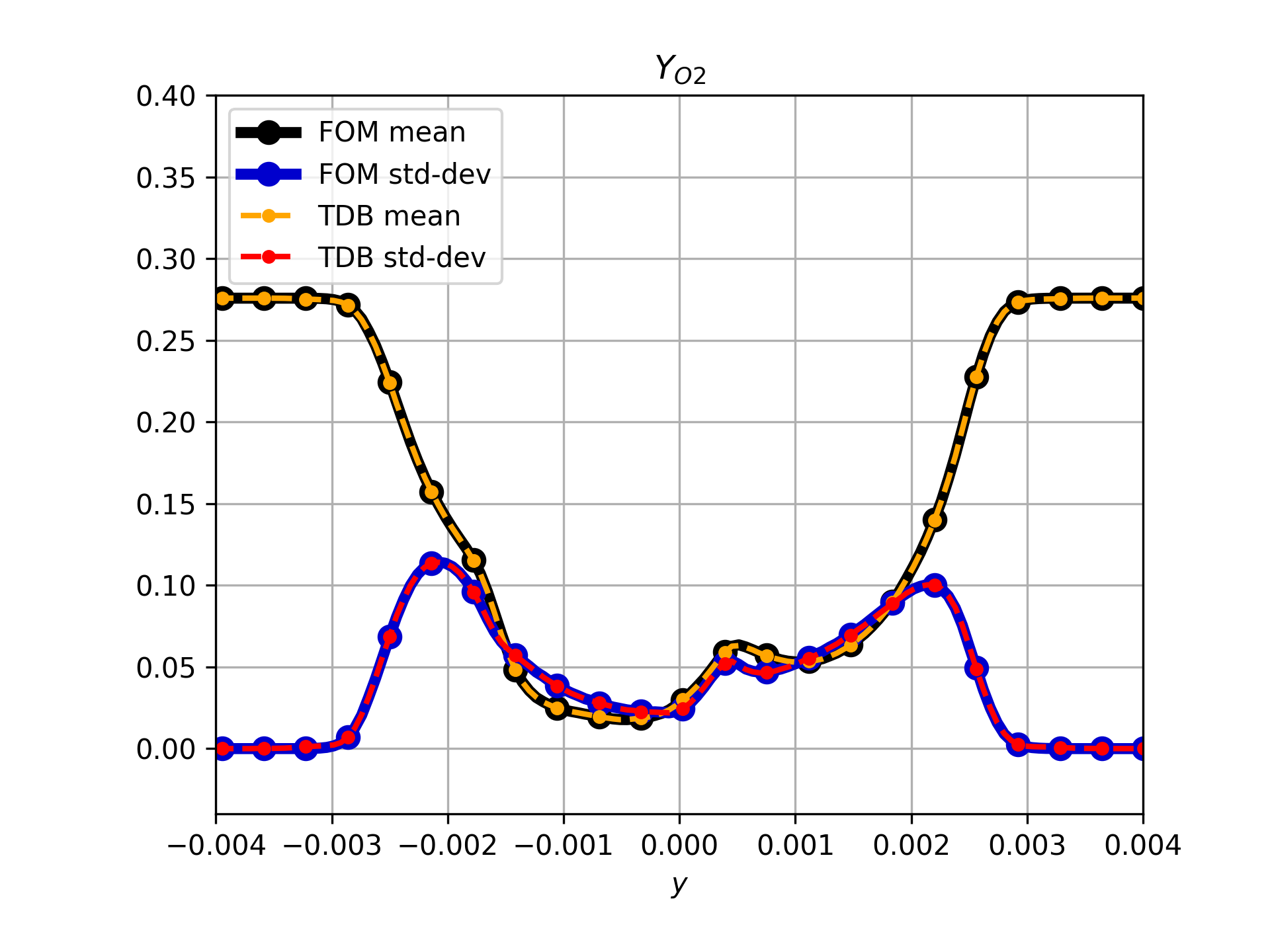}
        \caption{$t=40t_j$}
    \end{subfigure}
    \caption{Reynolds averages of $O_2$ (TDB with $r=6$).}
    \label{fig:RavO2}
\end{figure}

Simulations are conducted with $r=6$ and $r=8$. The singular values are shown on Fig.\ \ref{fig:SIGMA}. As expected the errors become smaller at the higher $r$ values (Fig.\ \ref{fig:ERRORS}). Moreover, the leading singular values of both low-rank approximations match very well. This indicates the leading singular values  have converged in both cases.
The fidelity of ROM  predictions is assessed via comparisons with FOM results. These comparisons are made for both instantaneous snap-shots and the Reynolds-averaged statistics.  
To show the overall flow structure, contour plots are provided of the $CO_2$ and $OH$ mass fractions in Figs.\ \ref{fig:CO2Field}-\ref{fig:OHField}. These contours show the formation of structures within the flow from the initial parallel layer. The overall compositional flow structure and the flow evolution are predicted well.  With $r=8$, the ROM predictions are almost identical to FOM results.  With $r=6$, the level of agreement is not as good.  In this case, the truncation  causes under- and overshoots in the instantaneous profiles  of the transport variables.  Here, the amplitude of oscillation is very low and does not cause any problems.  However, at  lower $r$ values, higher amplitudes of oscillations could lead to  serious errors.  For a more quantitative assessment, comparisons are made for the first and second Reynolds-moments of the mass fractions of several of the species.  These are shown in  Figs.\ \ref{fig:RavCO}-\ref{fig:RavO2} and indicate an   excellent performance of the ROM, especially for predicting the statistics of the major species.  

\begin{figure}
    \centering
    \begin{subfigure}[b]{0.49\textwidth}
        \centering
        \includegraphics[width=\textwidth]{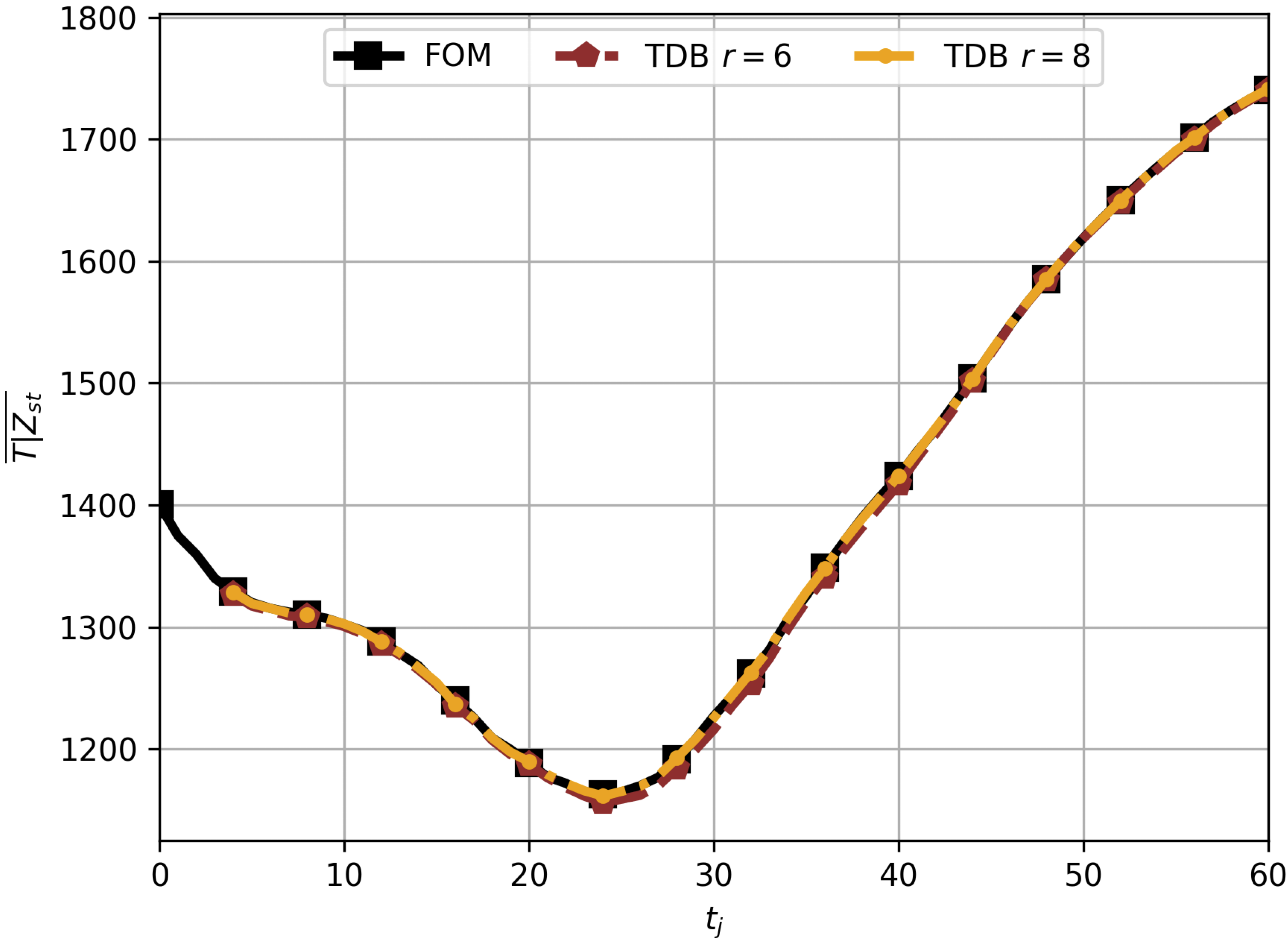}
        \caption{Full range}
    \end{subfigure}
    \begin{subfigure}[b]{0.49\textwidth}
        \centering
        \includegraphics[width=\textwidth]{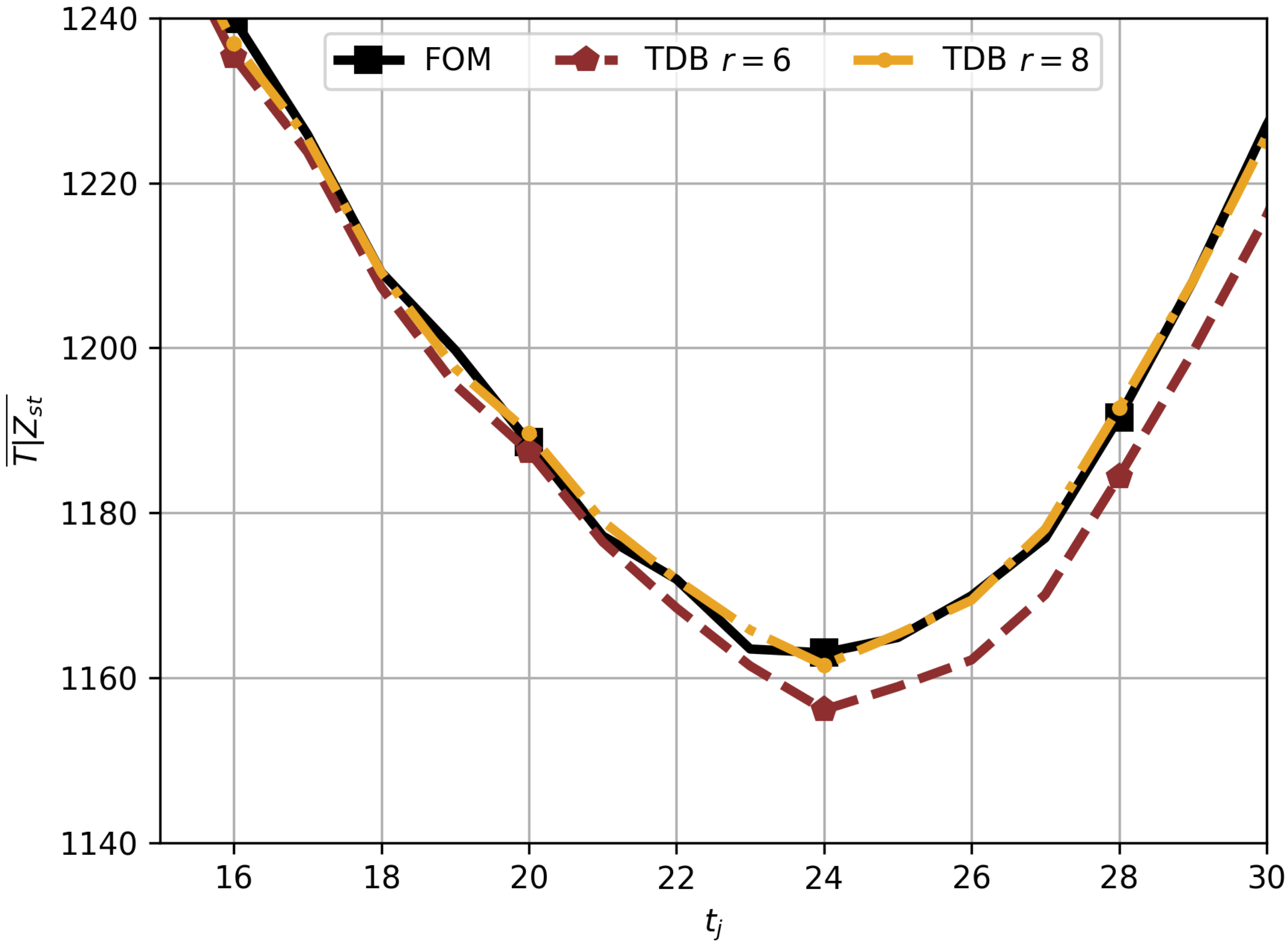}
        \caption{Zoomed-in $t \in [15t_j, 30t_j]$}
    \end{subfigure}
    \caption{Extinction-reignition demonstrated as a temporal evolution of conditionally averaged temperature at stoichiometric mixture fraction $Z_{st} \approx 0.42$.}
    \label{fig:TZst}
\end{figure}

The FDF-TDB solver is capable of capturing some of the complex non-equilibrium extinction/re-ignition effects as observed in this flame \cite{Hawkes2007Scalar}.
At initial times, when the mixture-fraction dissipation rates are large, the flame cannot be sustained and is locally extinguished.  At later times, when the dissipation values are lowered, the flame is re-ignited, and the temperature increases.  This dynamic is depicted in  Fig.\ \ref{fig:TZst}, where the expected temperature values conditioned on the mixture fraction are shown.  By $t=20t_j$ the temperature at the stoichiometric mixture fraction ($Z_{st} = 0.42$) decreases from $T = 1400 K$,  stays below extinction limit for a while, and then rises after $t \approx 25 t_j$. The agreement with FOM predictions is also very good for this conditional expected value.

\begin{figure}
    \centering
    \includegraphics[width=0.47\textwidth]{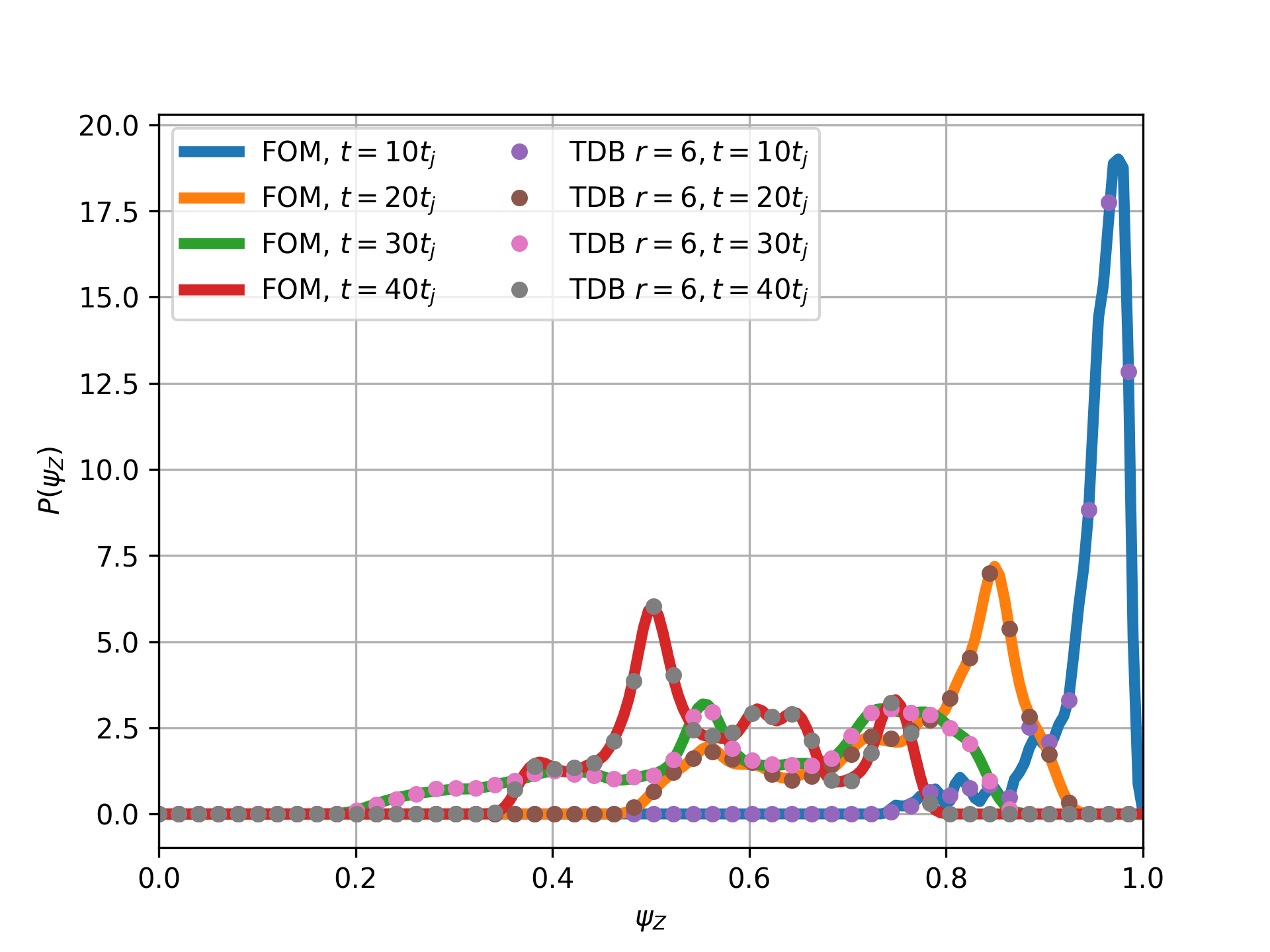}
    \caption{PDF of the mixture fraction $Z$ sampled along $y=0$ plane (line) at $t=10t_j$, $t=20t_j$, $t=30t_j$ and $t=40t_j$ for FOM and TDB with $r=6$.}
    \label{fig:PDF_Z}
\end{figure}

\begin{figure}
    \centering
    \begin{subfigure}[b]{0.49\textwidth}
        \centering
        \includegraphics[width=\textwidth]{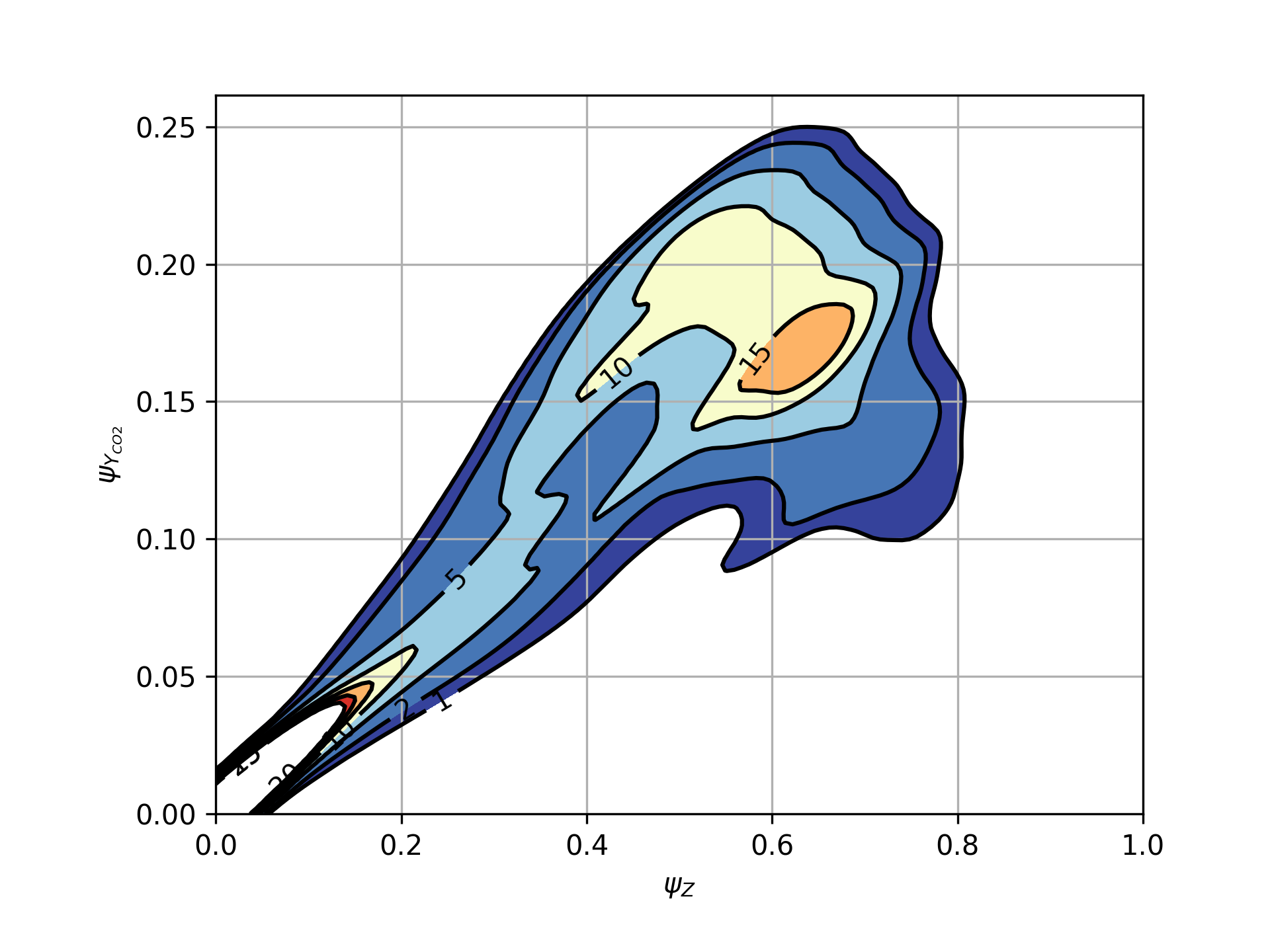}
        \caption{$r=6$}
    \end{subfigure}
    \begin{subfigure}[b]{0.49\textwidth}
        \centering
        \includegraphics[width=\textwidth]{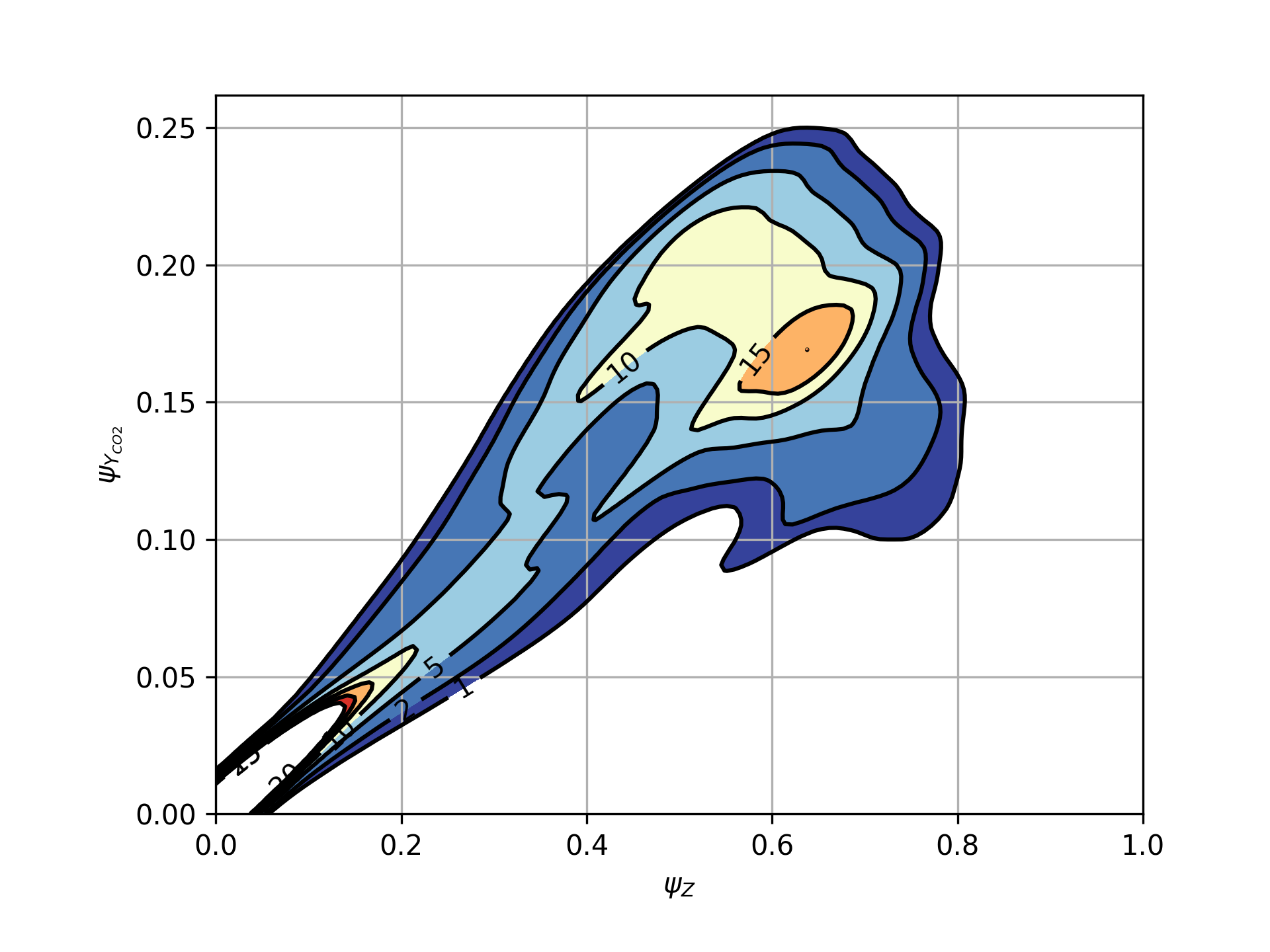}
        \caption{$r=8$}
    \end{subfigure}

    \begin{subfigure}[b]{0.49\textwidth}
        \centering
        \includegraphics[width=\textwidth]{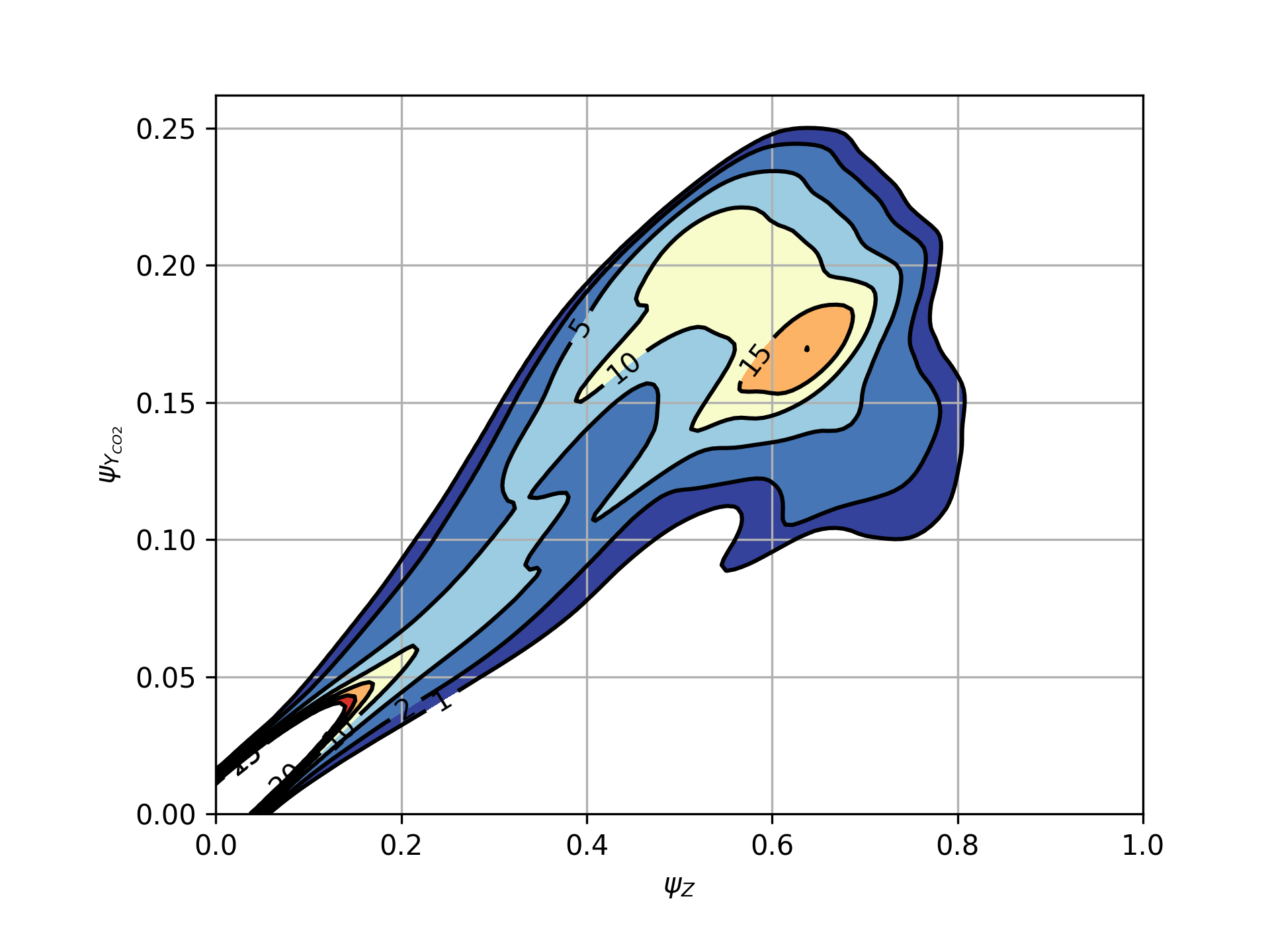}
        \caption{FOM}
    \end{subfigure}
    \caption{Joint PDF of $CO_2$ and $Z$ at $t=40t_j$.}
    \label{fig:JPDF_CO2}
\end{figure}

\begin{figure}
    \centering
    \begin{subfigure}[b]{0.49\textwidth}
        \centering
        \includegraphics[width=\textwidth]{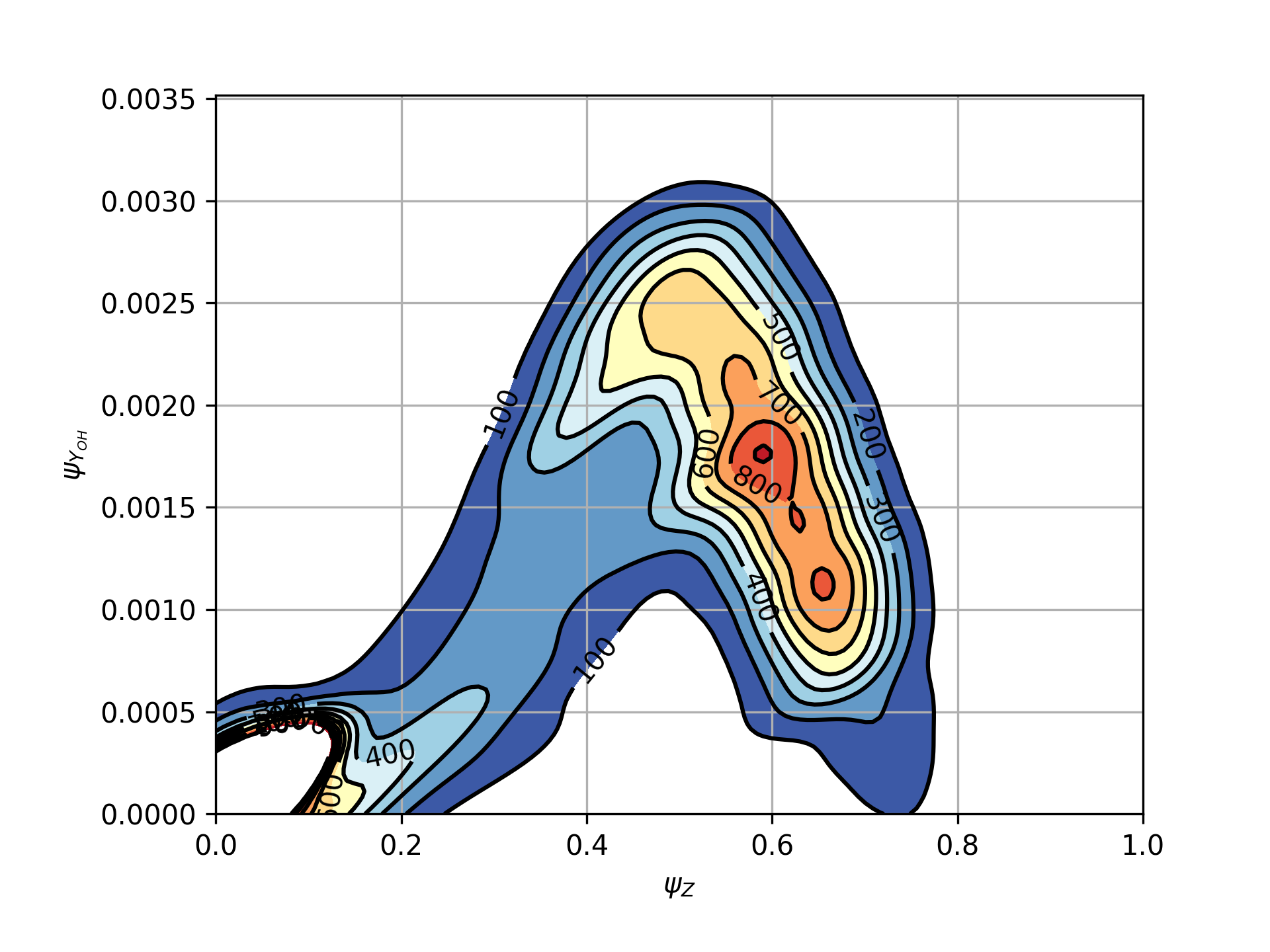}
        \caption{$r=6$}
    \end{subfigure}
    \begin{subfigure}[b]{0.49\textwidth}
        \centering
        \includegraphics[width=\textwidth]{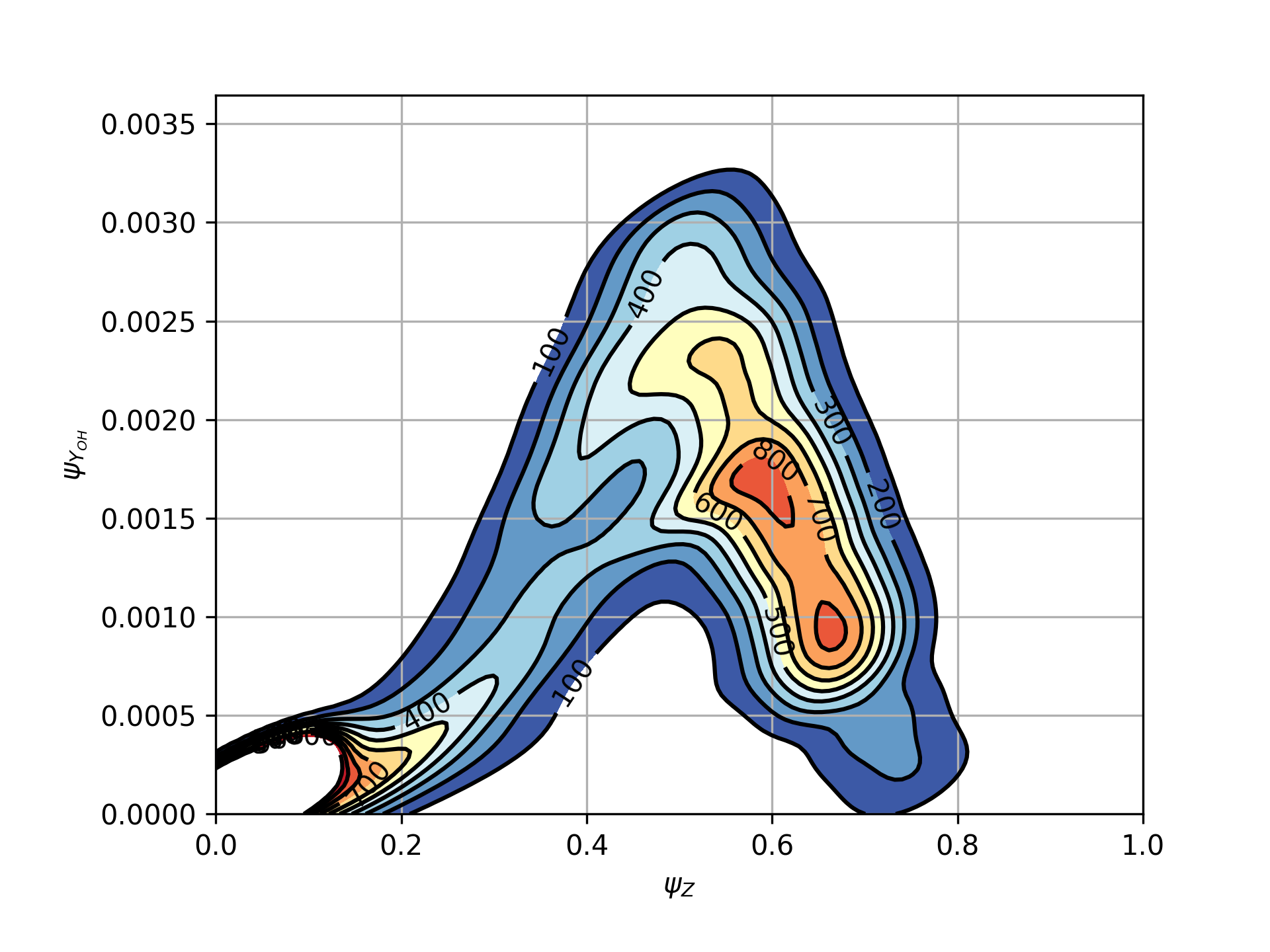}
        \caption{$r=8$}
    \end{subfigure}

    \begin{subfigure}[b]{0.49\textwidth}
        \centering
        \includegraphics[width=\textwidth]{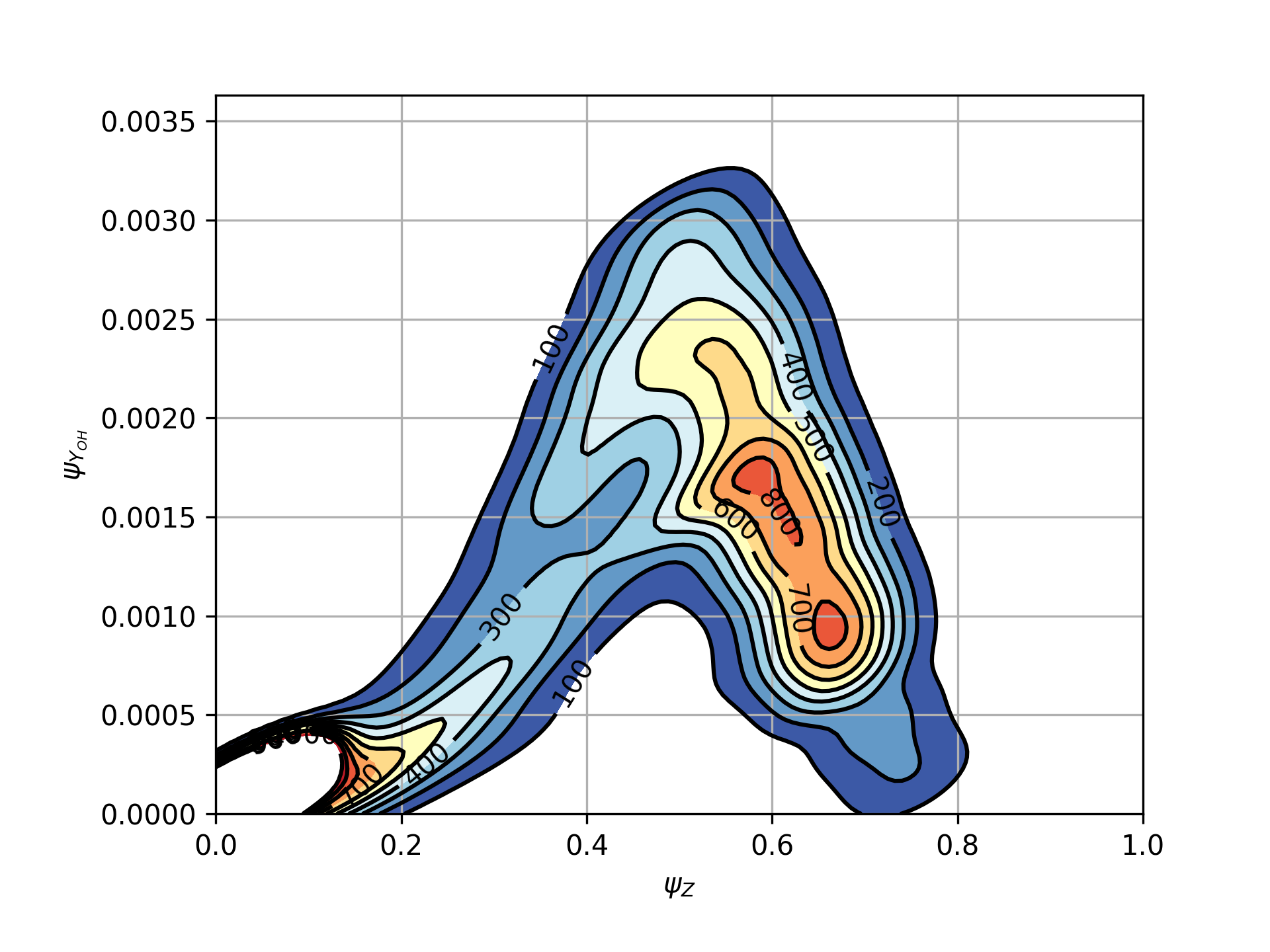}
        \caption{FOM}
    \end{subfigure}
    \caption{Joint PDF of $OH$ and $Z$ at $t=40t_j$.}
    \label{fig:JPDF_OH}
\end{figure}

A more comprehensive comparison between ROM and FOM results is made  by examination of the mixture fraction PDFs in  Fig.\ \ref{fig:PDF_Z}.   The  PDFs are generated by sampling of $N_{x} \times 2$ (2 cross-stream lines).  It is observed that even for $r=6$ the PDFs are approximated very well. The results for $r=8$ are nearly identical and are not shown.    To portray the dynamics of multi-scalar mixing and reaction, the joint PDFs of the scalar variables are considered. The joint PDFs of the mixture fraction and the mass fraction of the CO\textsubscript{2}, and those of the mixture fraction and mass fractions of $OH$ are shown in  Figs.\ \ref{fig:JPDF_CO2} and \ref{fig:JPDF_OH}, respectively.  In both cases, as the layers become fully mixed at $t=40t_j$, the PDFs tend to have a multivariate Gaussian-like distribution. In all cases, the ROM-predicted PDFs are in very good greements with those depicted by FOM. The predictions for $r=8$ are slightly better than those by $r=6$. 

\section{Summary and Conclusions}

Large eddy simulation via the filtered density function (FDF) has proven very effective for predictions of turbulence-combustion interactions.  The LES-FDF transport equation is most conveniently solved via Monte Carlo methods; such methods have shown significant success in predictions of a large number of complex turbulent combustion problems.  There is, however,  is a continuing need to reduce the computational cost of LES-FDF to make it more viable for applications to a broader class of turbulent combustion systems.  The present work makes progress in doing exactly so by developing and implementing a novel reduced-order model (ROM)  in which a lower  number of scalar variables are considered.  This ROM employs 
time-dependent bases (TDBs) and present  work is to develop an on-the-fly reduced-order model based on time-dependent bases (TDBs). 
The novelty of the methodology is it model-drivesn, operates in the fly and has the capability to capture the intricate dynamics of turbulence-chemistry interactions.  The FDF-TDB models is employed for   LES  of a CO/H\textsubscript{2} temporally developing jet flame. The results are assessed via detailed {\it a posteriori} comparative assessments against full-order LES-FDF of the same flame.  Excellent agreements are observed for the temporal evolution of all of the thermo-chemical variables.   The new methodology is shown to be particularly effective in capturing non-equilibrium turbulence-chemistry interactions.  This is demonstrated by capturing the flame-extinction and its re-ignition as observed via FDF-FOM and previous DNS.  The new LES-FDF-TDB simulator provides an excellent tool for affordable computational simulations of complex turbulent combustion systems.   Suggestions for future work: 

\begin{enumerate}

        \item The ROM as developed here is recommended for LES-FDF  with inclusions of more of the transport variables (e.g.\ velocity, pressure, $\cdots$). For that, the full self-contained formulation of the FDF transport \cite{NNGLP17} should be considered. 

    \item Resolution assessment in LES-FDF-TDB, like that in  all LES,  is of crucial importance \cite{DSMG07,NYSG10,SAMG20}, and is recommended.
    
    \item Significant recent developments have been made in fine-tuning  of the SGS models in FDF and its MC simulations \cite{ZGR2025}.  Implementation of these upgrades are recommended for FDF-TDB to facilitate  future applications of this powerful methodology for LES of a wider variety of complex  combustion systems.  

    \item The computational cost associated with the ROM transport equations can be reduced by implementation of the so-called CUR \cite{GTZ97} and/or deep neural networks (DNNs) as surrogate models \cite{Sheikhi2024,OE17} to model turbulent mixing or evaluate the chemical source term. To lower the flop costs of  ROM evolution equations, new methodologies that leverage oblique projection have been introduced \cite{NB23,DPNFB23}. Their implementation for future FDF-TDB would be very constructive. 

    \item  With the demonstration of its fidelity, the LES-FDF-TDB is expected to be employed for prediction  of  a wide variety of complex turbulent combustion systems.

\end{enumerate}

\section{Acknowledgments}
This work is sponsored by the US National Science Foundation (NSF) under Grant CBET-2152803 and by the Air Force Office of Scientific Research (PM: Dr.\ Fariba Fahroo) FA9550-25-1-0039.The computational resources are provided by the University of Pittsburgh Center for Research Computing.


\end{document}